\documentclass{SciPost}
\usepackage{amsmath}
\usepackage{amsfonts}
\usepackage{graphicx}
\usepackage{color}
\usepackage{verbatim}

\begin{document}

\newcommand{\beq}{\begin{equation}}
\newcommand{\eeq}{\end{equation}}
\newcommand{\beqa}{\begin{eqnarray}}
\newcommand{\eeqa}{\end{eqnarray}}
\newcommand{\ben}{\begin{enumerate}}
\newcommand{\een}{\end{enumerate}}
\newcommand{\hs}{\hspace{1.5mm}}
\newcommand{\vs}{\vspace{0.5cm}}
\newcommand{\note}[1]{{\color{red}  #1}}
\newcommand{\notea}[1]{{\bf #1}}
\newcommand{\new}[1]{{#1}}
\newcommand{\ket}[1]{|#1 \rangle}
\newcommand{\bra}[1]{\langle #1|}
\newcommand{\im}{\dot{\imath}}
\newcommand{\tg}[1]{\textcolor{blue}{#1}}
\newcommand{\f}{f^{\phantom{\dagger}}}

\begin{center}{\Large \textbf{
Spin Liquid versus Spin Orbit Coupling on the Triangular Lattice
}}\end{center}

\begin{center}
Jason Iaconis\textsuperscript{1*},
Chunxiao Liu\textsuperscript{1},
G\'abor B. Hal\'asz\textsuperscript{2},
Leon Balents \textsuperscript{2}
\end{center}

\begin{center}
{\bf 1} Department of Physics, University of California, Santa
Barbara, CA 93106-9530, USA\\
{\bf 2} Kavli Institute for Theoretical Physics, University of California, Santa Barbara, CA 93106-4030, USA\\
* jiaconis@physics.ucsb.edu
\end{center}

\begin{center}
\today
\end{center}

\section*{Abstract}
{\bf In this paper, we explore the relationship between strong spin-orbit coupling and spin liquid physics.  We study a very general model on the triangular lattice where spin-orbit coupling leads to the presence of highly anisotropic interactions.  We use variational Monte Carlo to study both $U(1)$ quantum spin liquid states and ordered ones, via the Gutzwiller projected fermion construction.  We thereby obtain the ground state phase diagram in this phase space.  We furthermore consider effects beyond the Gutzwiller wavefunctions for the spinon Fermi surface quantum spin liquid, which are of particular importance when spin-orbit coupling is present.
}

\section{Introduction}


Quantum spin liquids (QSLs) are exotic phases of correlated
electrons possessing highly entangled ground states, exotic
fractionalized excitations, and typically, the absence of  any
magnetic order \cite{citeulike:6827478, 0034-4885-80-1-016502}.
Historically, studies of QSLs focused on spin-rotationally invariant
Heisenberg models, but in recent years, strongly anisotropic
interactions arising from spin-orbit coupling have come under focus
\cite{SOCreview}.  In the famous Kitaev honeycomb model,
bond-dependent interactions lead to an exactly solvable model with a
spin liquid ground state \cite{Kitaev20062}. Remarkably, it was
later shown that these directional interactions can be generated in
real materials when spin-orbit effects are present
\cite{doi:10.1143/PTPS.160.155,PhysRevLett.102.017205}.  In turn, this has led to the recent
discoveries of many candidate `Kitaev' materials and has paved the
way for the study of spin liquid physics in spin-orbital systems.
One recent example of particular interest is the material
YbMgGaO$_4$ \cite{scirep, PhysRevLett.115.167203, nature,
PhysRevLett.117.097201, natphys}. This system very likely contains
directional interactions of significant strength. Moreover,
thermodynamic and inelastic neutron scattering measurements have
been interpreted as supporting a QSL state with a Fermi surface of
neutral spin-1/2 excitations, ``spinons'', in this material.

Spin-orbit generated interactions invariably lead to a strong
breaking of spin-rotation symmetry. A consideration of this symmetry
in spin liquids can then reveal new and unexpected physics. One
striking feature is that the lowered symmetry allows for new
distinct spin-liquid phases which do not exist in the rotationally
invariant case \cite{PhysRevB.86.085145, PhysRevB.90.174417}. There
exists a systematic method of classifying these phases, given by the
so-called projective symmetry group (PSG) \cite{PhysRevB.65.165113}.
This approach also gives a method for constructing a wave function
for each phase, as a Gutzwiller projection of a free fermion state.

We will study a very general spin-orbit coupled model on a
triangular lattice which is believed to describe YbMgGaO$_4$
\cite{PhysRevB.94.035107, PhysRevB.94.174424, 2016arXiv160806445L}
and focus specifically on the possibility that this model contains
spin liquid physics.  We look at the allowed spin liquid phases and
use the PSG as a starting point of our analysis.  However, our main
tool throughout this work is the variational Monte Carlo (VMC)
technique. With this numerical technique, one performs Monte Carlo
sampling of the quantum wave function in the many-body basis where
electrons are localized on each site, allowing one to
work with trial states which would otherwise be intractable.

In this paper, we broadly address three points.  First, we expound
on the relationship between our model and the PSG wave functions.
The VMC allows us to quantitatively compare the energies of the
different candidate QSL phases. This approach complements recent
studies that work with the states phenomenologically
\cite{2016arXiv161203447L, PhysRevB.96.075105}. We focus on gapless
spin liquids with emergent fermionic excitations and highlight the
differences between states with isolated Dirac-like quasiparticles
and those with a Fermi surface of gapless excitations.

Second, we compare the QSL states to magnetically ordered states,
seeking the region of stability of the former ones.  We show that a
QSL is favored if we allow for second-neighbor interactions, but
that spin-orbit effects work to reduce the size of this phase, in
agreement with Ref.~\cite{2017arXiv170302971Z}. We then go further
and show that, if a natural third-neighbor interaction is also
included, then the spin liquid phase is energetically competitive,
even in the presence of significant spin-orbit interactions.

Finally, we look at how spin-orbit coupling modifies the properties
of a QSL, and how this may lead to distinct observables for
experiment.  We develop a novel method to incorporate modifications
beyond the simplest Gutzwiller projected free fermion state into our
trial wave function.  This method proceeds by calculating many-body
corrections order by order in perturbation theory, and sampling
these using VMC.  We find that this technique is particularly useful
for our problem where spin-orbit interactions introduce qualitative
differences between the ground state and our trial states.  In
particular, we study the effect of spin-orbit coupling on the
energies of certain trial states and also demonstrate how unique
properties of these wave functions appear in the spin structure
factor and in thermal transport properties.

The remainder of this paper is structured as follows. In section II,
we define the general spin model on the triangular lattice that we
study in our work. In section III, we introduce the variational wave
functions given by the PSG analysis, which will form the basis for
the rest of our discussion. We first calculate the energies of the
different candidate spin liquid ans\"atze using variational Monte
Carlo, then allow for the possibility of magnetic order in our
simulation, and finally plot the full variational phase diagram for
our Hamiltonian. In section IV, we introduce our new method for
improving the simple PSG wave functions. We calculate the
corrections to the energy and the spin structure factor of the
spinon Fermi surface spin liquid state. We also show how the
spin-orbit interactions may result in an appreciable thermal Hall
conductivity in this system. Finally, in section V, we summarize our
results and discuss the relevance of our work to the material
YbMgGaO$_4$.

\section{The Model}

\begin{figure}[t]
\begin{center}
{\bf a)} \hspace{-2mm}
\includegraphics[scale=0.21]{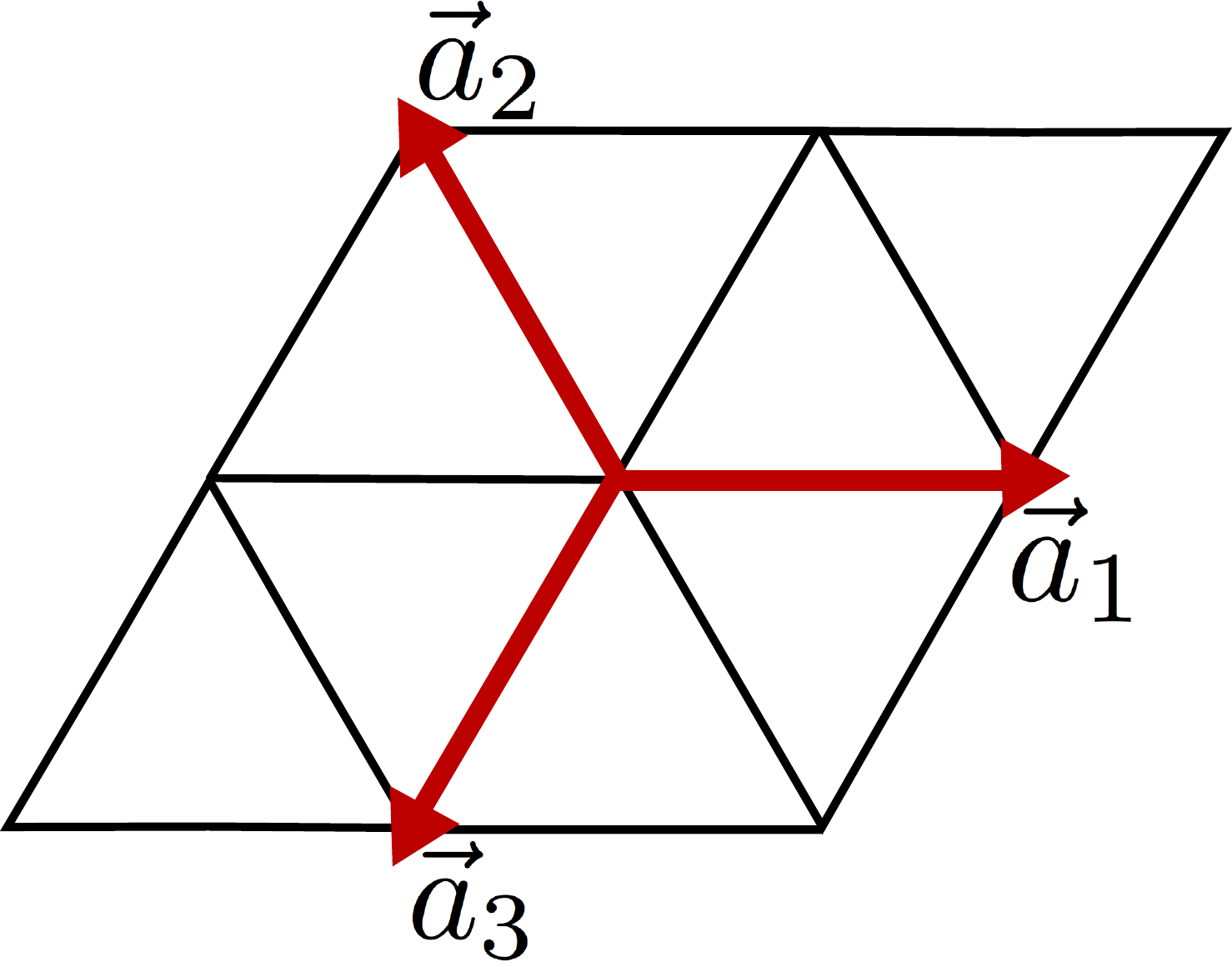} \hspace{5mm}
{\bf b)} \hspace{-1mm}
\includegraphics[scale=0.21]{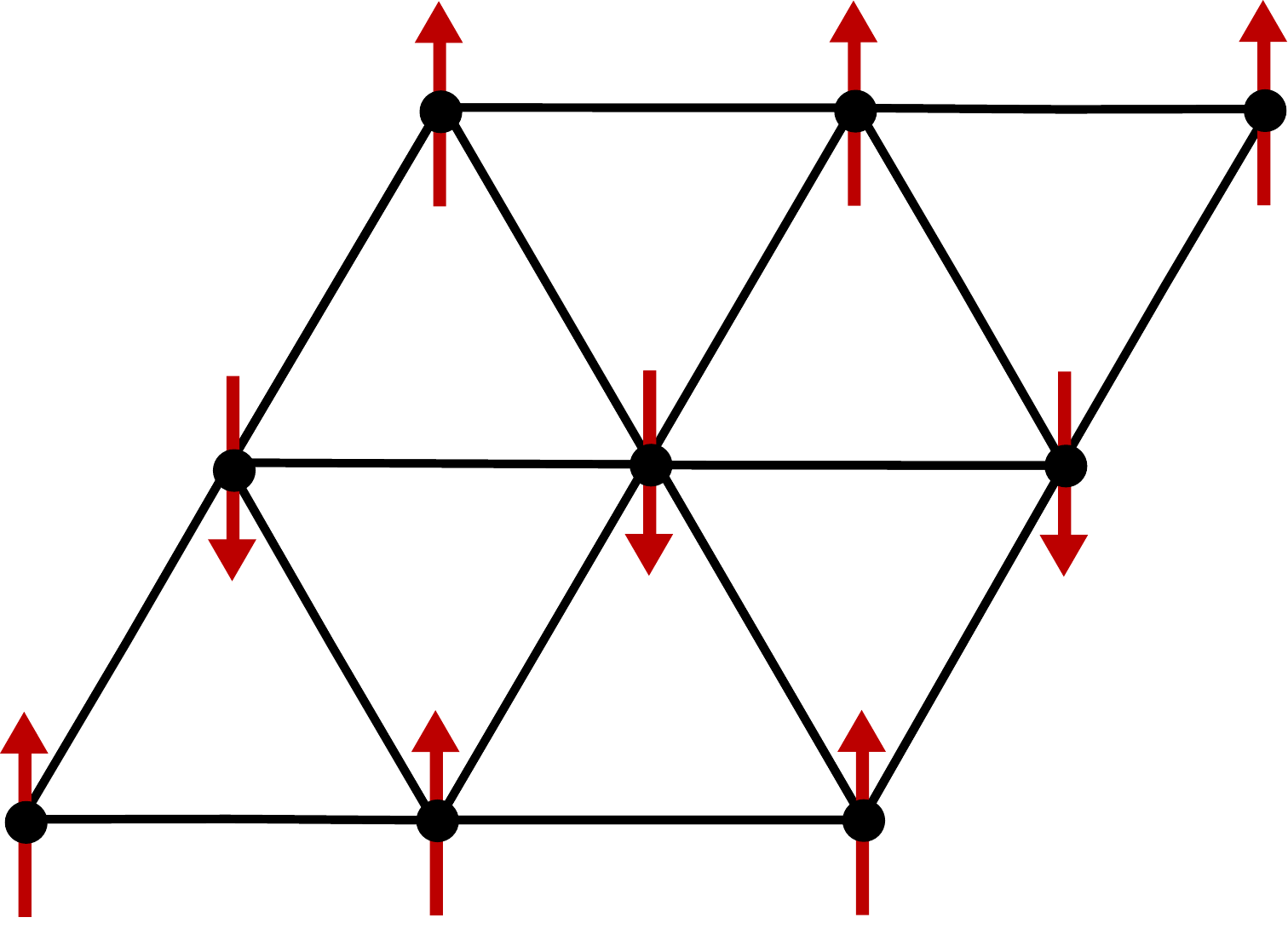} \hspace{5mm}
{\bf c)} \hspace{-1mm}
\includegraphics[scale=0.21]{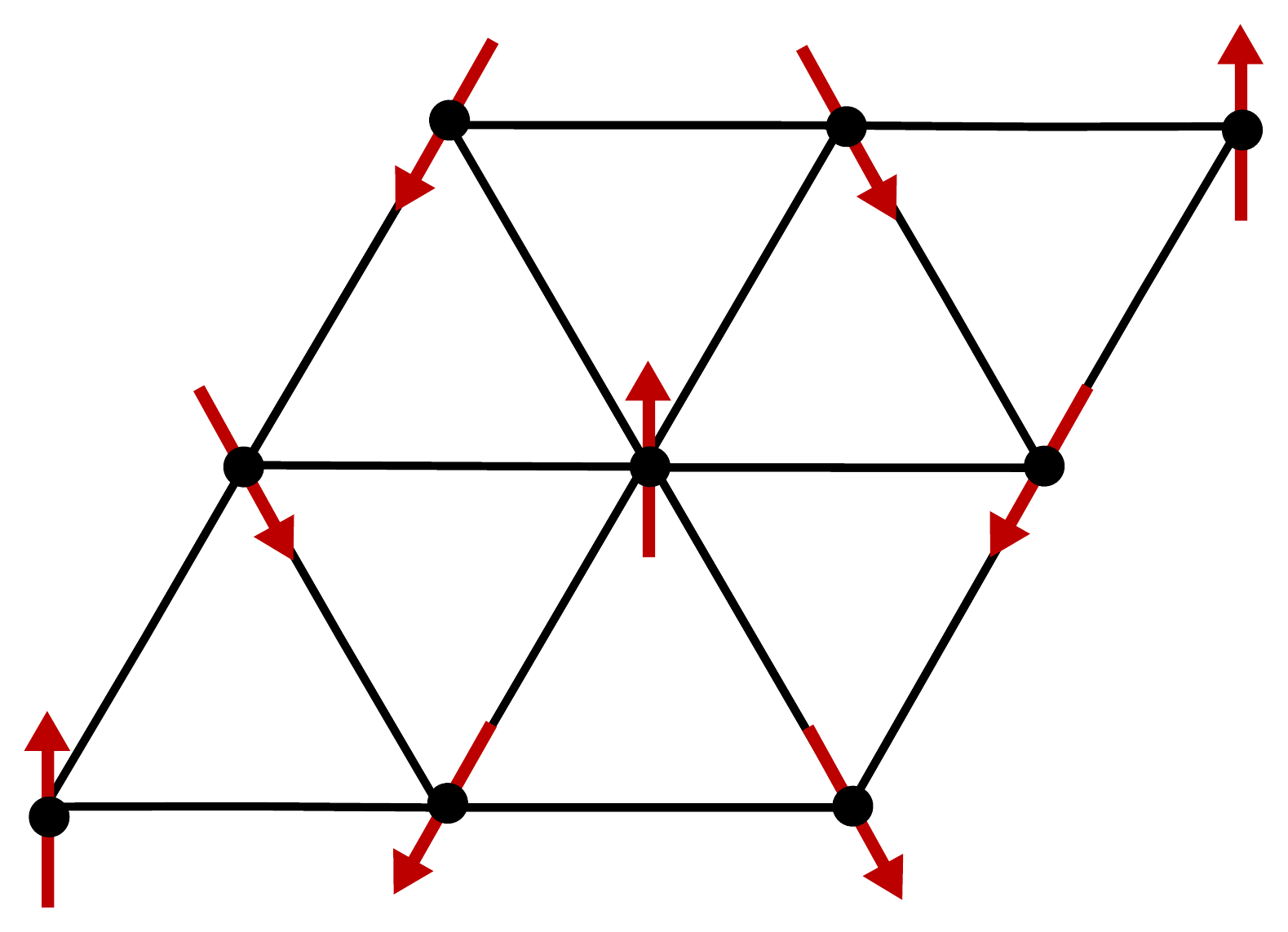}
 \caption{{\bf a)} The
three lattice bonds $a_1$, $a_2$, and $a_3$. The two commensurate
magnetic orders we consider are {\bf b)} stripe order and {\bf c)}
120$^\circ$
 antiferromagnetic
order.}
\label{fig:fig1}
\end{center}
\end{figure}

In many physical systems, the spin and orbital degrees of freedom of
the localized electrons are highly entangled.  In these cases, when
the rotation symmetry is broken by the surrounding crystal
structure, the spin-rotation symmetry is broken as well.
Superexchange processes then lead to the generation of highly
anisotropic terms in the effective spin Hamiltonian.  In these
strongly spin-orbit coupled systems, lattice symmetry
transformations are accompanied by an equivalent transformation in
spin space.  Following Ref.~\cite{PhysRevB.94.035107}, we consider
the Hamiltonian
\begin{eqnarray}
H &=& H_\pm + H_z + H_{\pm \pm} + H_{\pm z}, \nonumber \\ \nonumber \\
H_\pm &=& J_\pm {\sf H}_\pm = J_\pm \sum_{\langle i j \rangle} \left( S^+_i S^-_j \, + \, S^-_i S^+_j \right), \nonumber \\
H_z &=& J_z {\sf H}_z = J_z \sum_{\langle i j \rangle} S_i^z S_j^z, \label{Ham} \\
H_{\pm \pm} &=& J_{\pm \pm} {\sf H}_{\pm \pm} = J_{\pm \pm} \sum_{\langle i j \rangle} \left( \gamma_{ij} S^+_i S^+_j + \gamma_{ij}^* S^-_i S^-_j \right), \nonumber \\
H_{\pm z} &=& J_{\pm z} {\sf H}_{\pm z} \nonumber \\
&=& i J_{\pm z} \sum_{\langle i j \rangle} \left[ (\gamma_{ij}^*
S_i^z S_j^+ - \gamma_{ij}\, S_i^z S_j^-) + (i \leftrightarrow j)
\right], \nonumber
\end{eqnarray}
where $\gamma_{ij} = 1, e^{2\pi i/3}, e^{-2\pi i/3}$ for bonds
$\langle ij \rangle$ along the $\vec{a}_1, \vec{a}_2, \vec{a}_3$
directions, respectively (see Fig.~\ref{fig:fig1}a)). This is the
most general nearest-neighbor Hamiltonian which is invariant under
the symmetry generators of the system: the translations
$\mathcal{T}_{1,2}$ along the $\vec{a}_{1,2}$ directions, the
sixfold roto-reflection $\mathcal{S}_6$ within the plane of the
lattice, the twofold rotation $\mathcal{C}_2$ around a bond in the
$\vec{a}_3$ direction, and time reversal $\Theta$. (Note that the
threefold rotation $\mathcal{C}_3 = \mathcal{S}_6^2$ and the
inversion $\mathcal{I} = \mathcal{S}_6^3$ are both generated by the
sixfold roto-reflection. Conversely, the sixfold roto-reflection
$\mathcal{S}_6 = \mathcal{C}_3^2 \mathcal{I}$ is a combination of a
$120^\circ$ rotation and an inversion.) The symmetry generators are
all discrete and act simultaneously in real space and spin space. In
particular, they transform the coordinates $x_1, x_2$ of a general
lattice point $\vec{r} = x_1 \vec{a}_1 + x_2 \vec{a}_2$ as
\begin{eqnarray}
\mathcal{T}_1: && (x_1, x_2) \rightarrow (x_1+1, x_2), \nonumber \\
\mathcal{T}_2: && (x_1, x_2) \rightarrow (x_1, x_2+1), \nonumber \\
\mathcal{C}_2: && (x_1, x_2) \rightarrow (x_2, x_1), \\
\mathcal{S}_6: && (x_1, x_2) \rightarrow (x_1-x_2, x_1), \nonumber \\
\Theta: && (x_1, x_2) \rightarrow (x_1, x_2), \nonumber
\end{eqnarray}
while they transform the spin components $(S^x, S^y, S^z)$ as
\begin{eqnarray}
\mathcal{T}_{1,2}: && (S^x, S^y, S^z) \rightarrow (S^x, S^y, S^z), \nonumber \\
\mathcal{C}_2: && (S^x, S^y, S^z) \rightarrow (-\frac{1}{2} S^x + \frac{\sqrt{3}}{2} S^y, \frac{\sqrt{3}}{2} S^x + \frac{1}{2} S^y, -S^z), \\
\mathcal{S}_6: && (S^x, S^y, S^z) \rightarrow (-\frac{1}{2} S^x + \frac{\sqrt{3}}{2} S^y, -\frac{\sqrt{3}}{2} S^x - \frac{1}{2} S^y, S^z), \nonumber \\
\Theta: && (S^x, S^y, S^z) \rightarrow (-S^x, -S^y, -S^z). \nonumber
\end{eqnarray}
Importantly, the Hamiltonian does not generically have a continuous
spin-rotation symmetry because the XXZ terms ${\sf H}_{\pm}$ and
${\sf H}_z$ break the $SU(2)$ spin symmetry down to an in-plane
$U(1)$ spin symmetry, while the remaining terms ${\sf H}_{\pm \pm}$
and ${\sf H}_{\pm z}$ further break the $U(1)$ spin symmetry down to
discrete spin symmetries that are intertwined with appropriate
lattice symmetries.

It is helpful to write the ${\sf H}_{\pm\pm}$ and ${\sf H}_{\pm z}$
terms in a slightly different form to further expose the symmetries:
\begin{eqnarray}
{\sf H}_{\pm\pm} &=& \sum_{\langle ij \rangle} (\gamma_{ij} S^+_i S^+_j + \text{h.c.}) \nonumber \\
&=& 4\sum_{\langle ij \rangle} \left[ (\vec{S}_i \cdot
\hat{n}_{ij})(\vec{S}_j \cdot \hat{n}_{ij}) - \frac{1}{2}(S_i^x
S_j^x + S_i^y S_j^y) \right], \nonumber \\
{\sf H}_{\pm z} &=& \sum_{\langle ij \rangle} \left[ (i \gamma_{ij} S^+_i S^z_j + \text{h.c.}) + (i \leftrightarrow j) \right] \\
&=& 2\sum_{\langle ij \rangle} \left[ \{ (\vec{S}_i \times
\hat{n}_{ij}) \cdot \hat{z} \} \, S_j^z + (i \leftrightarrow j)
\right]. \nonumber
\end{eqnarray}
where $\hat{n}_{ij}$ is the unit vector pointing from site $i$ to site $j$.
The term ${\sf H}_{\pm\pm}$ has a `clock' structure where spins
would like to align along the 120$^\circ$ bond directions, and the
term ${\sf H}_{\pm z}$ also has a bond dependent structure that
incorporates the $\hat{z}$ direction.

There are several cursory reasons one may expect to find spin liquid
physics in this model. For one, due to its strong frustration, the
triangular lattice has a long and storied history as a spin liquid
candidate \cite{PhysRevB.72.045105, PhysRevB.73.155115,
PhysRevB.93.144411, PhysRevB.92.041105, PhysRevB.92.140403}. Beyond
that, the form of the anisotropic part of $H$ is highly reminiscent
of the interactions in the Kitaev honeycomb model
\cite{Kitaev20062}, where the direction-dependent spin-spin
interactions frustrate the coupling in a way which renders all
magnetic orders energetically unfavorable.

\section{Spin Liquid Wave functions}

\subsection{Generalities of parton wavefunctions}
\label{sec:gener-part-wavef}

The ground state wave function in a quantum spin liquid is
completely symmetric under all the symmetries of the Hamiltonian.
The PSG gives a systematic classification of the allowed spin liquid
phases under such a set of symmetries \cite{PhysRevB.65.165113}. In
the process, it also gives a construction of a representative wave
function for each phase. It is a surprising fact that, in many
cases, the number of allowed spin liquid phases increases as the
symmetry is reduced \cite{PhysRevB.90.174417,PhysRevB.86.085145}.
Spin liquids are fundamentally defined by their fractionalized
quasiparticle excitations, whose behavior can be described
phenomenologically by a mean-field Hamiltonian.  The PSG classifies
the fractionalized symmetry by identifying the allowed form of the
mean-field Hamiltonians.  In general, these excitations can realize
the symmetries of the original Hamiltonian in a nontrivial manner.

One starts by writing the physical spin operator $\vec{S}_i$ in terms of fermionic parton operators:
\begin{eqnarray}
\vec{S}_i = \frac{1}{2} f_{i \alpha}^\dagger \vec{\sigma}_{\alpha \beta} \f_{i \beta}.
\end{eqnarray}
The parton operators $f_{i \sigma}^{\vphantom\dagger},
f_{i\sigma}^\dagger$ live in a larger Hilbert space than the spins
$S_i$.  To remedy this, one must also include the strict gauge
constraint on the allowed states:
\begin{eqnarray}
\sum_\sigma f_{i \sigma}^\dagger \f_{i \sigma} = 1.
\label{constraint}
\end{eqnarray}
In this paper, we enforce Eq.~(\ref{constraint}) at the level of the
wave function. This is accomplished by applying the Gutzwiller
projection operator $\mathcal{P}$ to a state $\ket{\psi_0}$ in the
fermionic space:
\begin{eqnarray}
\ket{\Psi} &=& \mathcal{P} \ket{\psi_0} , \nonumber \\
\hspace{5mm} \mathcal{P} &=& \prod_i n_i(2-n_i). \label{gutz1}
\end{eqnarray}
The projected wave function $\ket{\Psi}$ lives in the proper Hilbert
space of spins and, with a suitable choice of $\ket{\psi_0}$, is
highly entangled in real space. Furthermore, with some minor
improvements, such an ansatz can be made to give variational
energies which are competitive with the most state of the art 2D
DMRG calculations \cite{PhysRevB.93.144411}.

For the state $\ket{\psi_0}$, we take a ``mean field'' wavefunction,
which is the ground state of some quadratic fermion Hamiltonian. The
parameters of that fiduciary Hamiltonian then become variational
parameters in the ansatz.  When the fermions are allowed to hop in
the mean field Hamiltonian, the partons become deconfined in the
corresponding spin liquid phase. In general, the quadratic
mean-field Hamiltonian can be written as
\begin{eqnarray}
\mathcal{H}_{mf} &=& \sum_{i,j,\alpha} \text{Tr} \left[\sigma^\alpha \Phi_i u_{ij}^\alpha \Phi_j^\dagger \right], \\
\Phi_i &=& \left( \begin{array}{cc}
\f_{i\uparrow} & f_{i\downarrow}^\dagger \\
\f_{i\downarrow} & -f_{i\uparrow}^\dagger
\end{array} \right),
\end{eqnarray}
where $\alpha = 0,x,y,z$. A local gauge transformation, such as
$f_{i\sigma} \rightarrow e^{i\theta_i \sigma^z}f_{i\sigma}  $,
changes $H_{mf}$ but leaves the physical spin operator $\vec{S}_i$
unchanged. Since the physical wave function is unchanged, all
mean-field Hamiltonians related by such local gauge transformations
must be equivalent. The parton Hamiltonian $\mathcal{H}_{mf}$ can
therefore ostensibly break the symmetries of $H$ as long as there
exists a local gauge transformation which restores the symmetry. In
this case, we say that the quasiparticle realizes the symmetry
nontrivially. The role of the PSG is to determine the set of allowed
mean-field Hamiltonians which cannot be connected to each other by
such a gauge transformation. Importantly, $\mathcal{H}_{mf}$ is
always invariant under some global transformations $\Phi \rightarrow
\Phi \cdot W$, where $W \in G$. The group $G \supseteq \mathbb{Z}_2$
of such global transformations is known as the `invariant gauge
group' (IGG) and determines the form of the gauge group around which
fluctuations of the gauge field may occur. In this work, we consider
$U(1)$ spin liquids with IGG = $U(1)$.

A more complete study would also include $\mathbb{Z}_2$ spin liquids
(IGG = $\mathbb{Z}_2$).  However, even restricting to
nearest-neighbor couplings, there are at least 18 different
$\mathbb{Z}_2$ mean-field ans\"atze.  To avoid this complexity, we
neglect these candidate QSLs for the present work.  This is at least
consistent with recent work on several related triangular lattice
spin systems, for which the $U(1)$ spin liquids have proven to have
competitive energies \cite{PhysRevB.72.045105,PhysRevB.93.144411}.
Furthermore, the spinon Fermi surface QSL suggested by several
previous papers for YbMgGaO$_4$ falls into the $U(1)$ class.

\subsection{Six specific parton states}
\label{sec:six-specific-parton}

The PSG classification of $U(1)$ QSLs for the space group of our
model was done in Ref.~\cite{2016arXiv161203447L}.  There are 6
distinct nearest-neighbor mean-field Hamiltonians:
\begin{align}
& \mathcal{H}_{mf}^{(1)} = \sum_{\langle i j \rangle, \sigma}
\left[ t_{ij}^{\phantom\dagger} f_{i\sigma}^\dagger \f_{j \sigma} + \text{h.c.} \right], \hspace{13mm} \tag{A1/B1} \\
& \mathcal{H}_{mf}^{(2)}  = i \sum_{\langle i j \rangle} \left[
t_{ij}^{\phantom\dagger} f_{i \alpha}^\dagger (\vec{\sigma}_{\alpha
\beta} \cdot \vec{n}_{ij}) \f_{j \beta} + \text{h.c.} \right] ,\hspace{2mm} \tag{A2/B2}\\
& \mathcal{H}_{mf}^{(3)} = i \sum_{\langle i j \rangle} \Big[
t_{ij}^{\phantom\dagger} f_{i \alpha}^\dagger \{
(\vec{\sigma}_{\alpha \beta} \times \vec{n}_{ij})\cdot \hat{z} \}
\f_{j \beta} + \lambda_{ij}^{\phantom\dagger} f_{i
\alpha}^\dagger \sigma^z_{\alpha \beta} \f_{j\beta} + \text{h.c.}
\Big]. \tag{A3/B3} \end{align}
The ground state of each mean-field Hamiltonian defines
$\ket{\psi_0}$ for the corresponding type of QSL.  We distinguish
two versions for each mean-field Hamiltonian
$\mathcal{H}_{mf}^{(n)}$, which differ only in the way translation
symmetry is realized.  In the A states, translation acts in the
usual way as $t_{ij} = -1$ for all nearest-neighbor bonds $\langle ij
\rangle$.  Conversely, in the B states, translation acts
nontrivially; this is achieved by setting $t_{ij} = \pm 1$ such that
the unit cell is doubled and a $\pi$ flux is thread through every
other triangle. In the A1/B1/A2/B2 cases, there are no variational
parameters (since the overall scale of the Hamiltonian leaves its
ground state unchanged), while in the A3/B3 cases, there is a single
variational parameter $\lambda/t$.

We note that, importantly, the spinon band structure determines the
physical properties of the wave functions and that it is gapless in
all 6 states. This is necessary for a U(1) spin liquid to be stable
in two dimensions. We now discuss some aspects of these states.

The (A1) state has no mixing between the up and down spin states. In
order to satisfy the constraint $ \langle f_i^\dagger \f_i \rangle =
1$, the band structure then must contain a large Fermi surface. We
refer to this state as the {\bf uniform Fermi surface (uFS)} or {\bf
spinon metal} state.  Notably, although the microscopic Hamiltonian
$H$ has only discrete symmetries, the mean-field Hamiltonian of this
uFS state is spin-rotationally invariant. This accidental ``emergent
SU(2) symmetry'' is surprisingly robust, and is not an accident of
assuming a nearest-neighbor form for $\mathcal{H}_{mf}$. In fact,
the PSG does not allow any spin-dependent terms [which would break
SU(2) symmetry] in $\mathcal{H}_{mf}$, even for hoppings of
arbitrary distance. The argument for this hinges on the fact that
both time-reversal ($\Theta$) and inversion ($\mathcal{I}$)
symmetries act trivially in this class. First, the operators which
implement these symmetries both involve a complex conjugation,
time-reversal by definition and inversion due to a site-exchange
which corresponds to a Hermitian conjugation. Then, since spin is
even under inversion and odd under time reversal, it is odd under
their combination, and so a spin-dependent term with any complex
coefficient is forbidden in the presence of such a combined
symmetry.

The (B1) state also has no mixing of the spin states, but
translations act nontrivially on the spinons. The unit cell is then
doubled and the band structure contains two Dirac cones. We
therefore refer to this state as the {\bf Dirac spin liquid} state.
The uFS and Dirac states are the two $U(1)$ spin liquids that can
also occur in rotationally invariant systems. Note, however, that
spin-dependent quadratic terms are not generically prohibited in the
case of the (B1) state and that they in fact appear at the level of
third-nearest-neighbor hoppings.

The (A2) and (B2) states are called the {\bf 120$^\circ$ clock spin
liquid (ClSL)} and the {\bf 120$^\circ$ clock + $\pi$ spin liquid
(Cl$\pi$SL)}, respectively. These ans\"atze do mix the spin flavors
and orbital degrees of freedom by including bond dependent hoppings.
The band structures in both cases contain protected Dirac cones at
the $\Gamma$, $M$, and $K$ points in the Brillouin zone.

The (A3) state, called the {\bf Rashba spin liquid (RSL)}, also has
Dirac cones at the $\Gamma$, $M$, and $K$ points when $\lambda=0$ or
$t=0$, and a gap opens at the $\Gamma$ point for intermediate values
of $\lambda/t$. Finally, the (B3) state, called the {\bf Rashba +
$\pi$ spin liquid (R$\pi$SL)}, contains 4 bands and a small Fermi
surface for intermediate values of $\lambda/t$.

\subsection{Energetics of PSG Wave Functions}

The PSG method gives us the full set of allowed free fermion wave
functions that are invariant under the symmetries of the system once
the gauge constraint, Eq.~(\ref{constraint}), is enforced. It tells
us nothing, however, about the energies of these wave functions.
The PSG gives us a starting ansatz, but is completely agnostic about
which state may actually be the ground state.

One simple way to proceed is to work directly with the single
particle wave functions by satisfying Eq.~(\ref{constraint}) on
average: $\frac{1}{N}\sum_{i, \sigma} \langle f_{i\sigma}^\dagger
\f_{i \sigma}\rangle = 1$. However, such a mean field approach
requires an infinite number of approximations, the resulting wave
functions do not even live in the proper Hilbert space, and thus it
cannot give reliable energy estimates.  Instead, we carry out a
variational analysis based on the fully projected wavefunctions in
Eq.~\eqref{gutz1}.   We calculate the variational energy $E_s =
\langle \Psi_s|H|\Psi_s\rangle$, where $s$ indicates one of the six
QSL ans\"atze.

\begin{figure}[t]
\begin{center}
{\bf a)} \hspace{-4mm}
\includegraphics[scale=0.084]{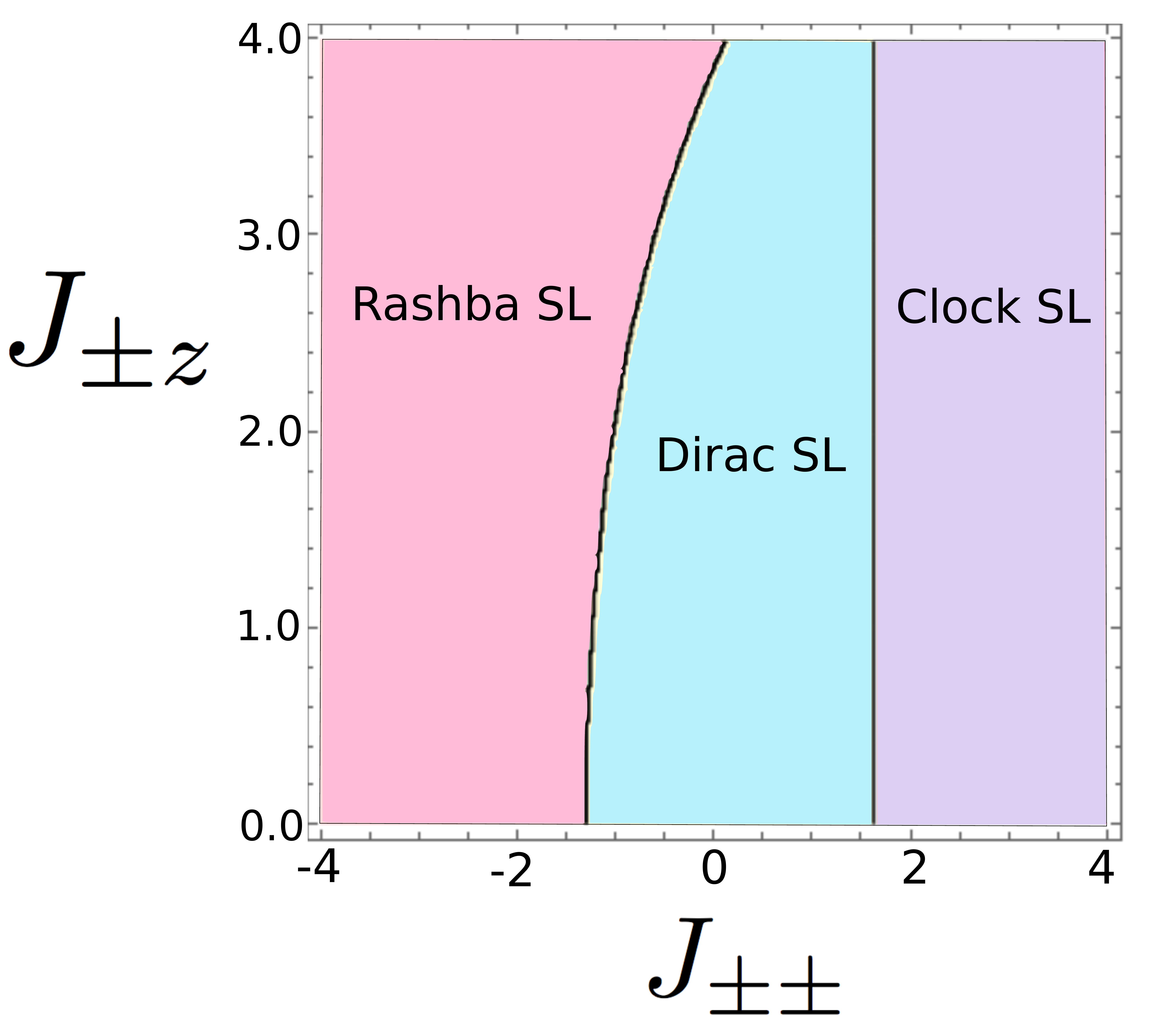}
\hspace{15mm} {\bf b)} \includegraphics[scale=0.085]{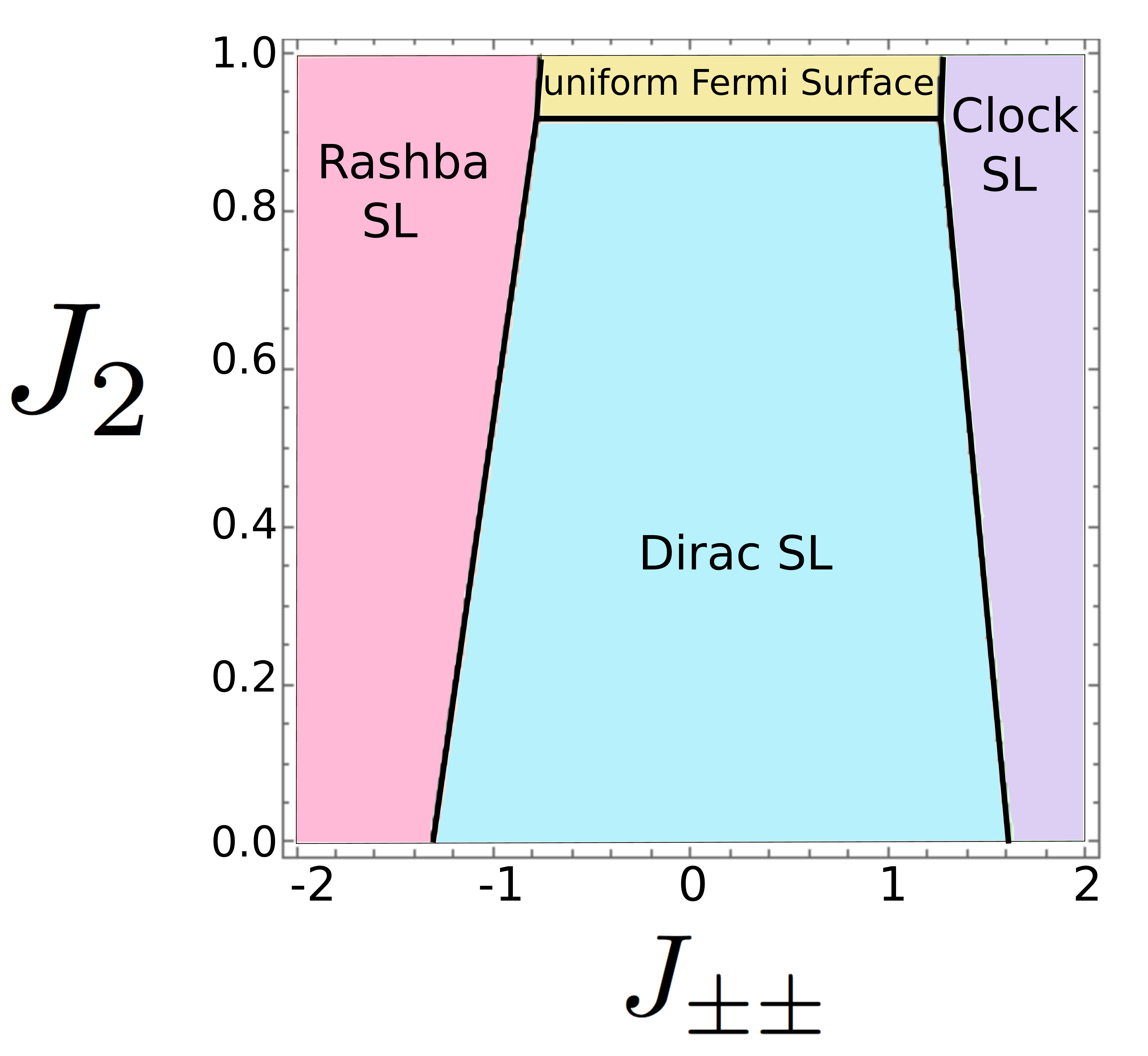} \caption{
The phase diagram showing only the lowest-energy spin liquid ground
states {\bf a)} in the $J_{\pm\pm}-J_{\pm z}$ plane with $J_2=0$,
and {\bf b)} in the $J_{\pm\pm}-J_2$ plane when $J_{\pm z}=0$.
We set the third neighbor coupling $J_3 = 0$.
All energies are measured in units of $J_{\pm} = 1$.
See the main text for a
description of the further neighbor terms $J_2$ and $J_3$.}
\label{fig:fig2}
\end{center}
\end{figure}

The results are highly constrained by how the projective symmetries
are implemented in the given mean-field Hamiltonian. In particular,
the uniform Fermi surface and Dirac spin liquid states are
completely $SU(2)$ invariant, and therefore the expectation values
of the $J_{\pm\pm}$ and $J_{z\pm}$ terms vanish in these states.
Similarly, while both the `clock' and `Rashba' Hamiltonians have
some spin-orbit terms, only the Rashba Hamiltonians include
spin-orbit terms both within and perpendicular to the $xy$ plane.
Consequently, the `clock' wave functions also yield vanishing
expectation values for the $J_{\pm z}$ terms.

We performed a variational Monte Carlo simulation and measured the
energies of each of our trial wave functions on finite size lattices
for system sizes up to $N=32\times32$ sites. Each mean-field wave
function, when projected, gives a different pattern of entangled
spins, giving rise to different spin correlations. When $\lambda=0$,
none of the wave functions have any free parameters. Setting
$J_{\pm} = 1$ and scaling to the thermodynamic limit, the
corresponding energy densities are then given by
\begin{eqnarray}
E_{Dirac}/N &=& -0.7050(1)[1+J_z/4], \hspace{10mm} \nonumber\\
E_{uFS}/N &=& -0.4682(5)[1+J_z/4], \hspace{-20mm} \\
E_{Clock}/N &=& -0.0645(2) + 0.325(1) J_z - 0.716(1) J_{\pm\pm}, \nonumber\\
E_{Rashba}/N &=& -0.1663(4) + 0.258(1) J_z + 0.741(1) J_{\pm\pm}, \nonumber\\
E_{Cl\pi}/N &=& -0.0619(6) - 0.321(1) J_z - 0.582(1) J_{\pm\pm}, \nonumber\\
E_{Rsh\pi}/N &=& +0.1173(4) + 0.256(1) J_z + 0.525(1) J_{\pm\pm}.
\nonumber
\end{eqnarray}
A few observations are apparent.  First, we see that the (Cl$\pi$SL)
and (R$\pi$SL) ans\"atze are never competitive energetically in our
regimes of interest.  While the Dirac state has the lowest energy at
$J_{\pm\pm}=0$, the clock and Rashba spin liquid states become
energetically favorable for large positive and negative
$J_{\pm\pm}$, respectively. The Rashba states (and only the Rashba
states) have an energy which is modified by including $\lambda \neq
0$, which is beneficial only when $J_{\pm z}\neq 0$. In this case,
we determine the optimal Rashba state for a given value of $J_{\pm
z}$ by numerically minimizing the energy with respect to
$\lambda/t$.

The results for a full comparison of energies are presented in
Fig.~\ref{fig:fig2}a), which shows the state of lowest variational
energy amongst the 6 QSLs for all values of $J_{\pm \pm}$ and
$J_{\pm z}$. (Note that the phase diagram is qualitatively similar
for all values of $J_z$.) Looking ahead, it has been suggested
\cite{natphys} that next-nearest-neighbor interactions may be
important in stabilizing a spin liquid ground state for our model.
We therefore also looked at the variational energies of our
ans\"atze when XXZ-like next-nearest-neighbor interactions are
added (see Eq.~\ref{Ham2} in Sec. 3.4.2). In Fig.~\ref{fig:fig2}b), we plot the lowest energy states as
a function of the next-nearest-neighbor coupling $J_2$ for $J_{\pm
z}=0$. Notice that the Fermi surface state only becomes competitive
in energy for very large next-nearest-neighbor coupling.

\subsection{Magnetic Order}

\begin{figure*}[t]
\begin{center}
\hspace{-7mm} \includegraphics[scale=0.2525]{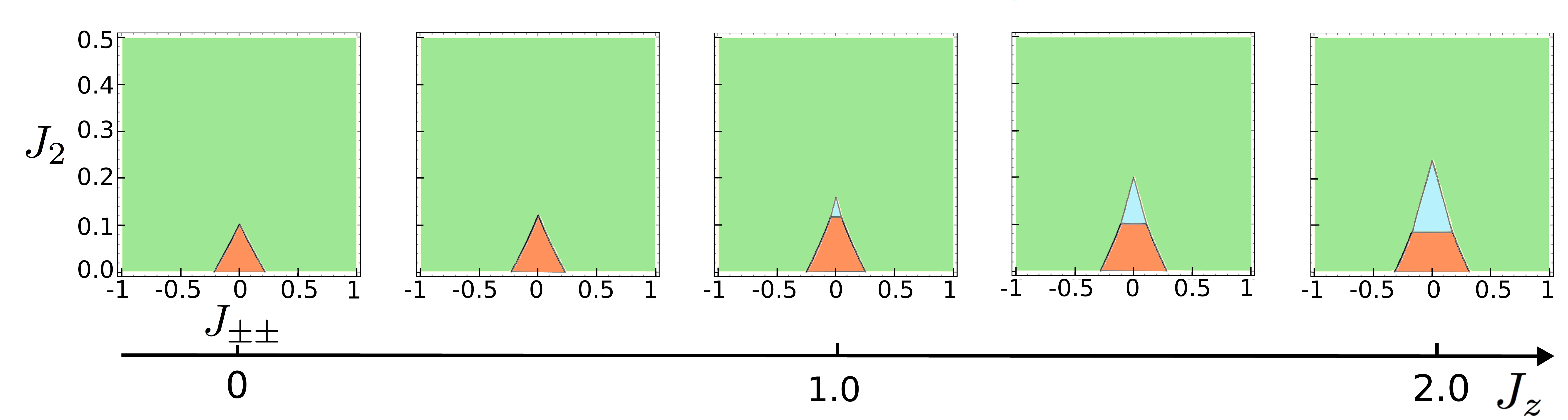}
\caption{The full $J_{\pm\pm}-J_2-J_z$ magnetic phase diagram for
$J_3 = J_{\pm z} = 0$. Green is stripe order, red is 120$^\circ$ AFM
order, and blue is the Dirac spin liquid phase.  Stripe order
dominates the phase diagram, except for small $J_2$ and
$J_{\pm\pm}$. The spin liquid regime also depends strongly on the
value of $J_z$ and is greatly reduced when $J_z$ moves away from the
isotropic point $J_z=2$. The horizontal axis on each subplot gives
the value of $J_{\pm\pm}$. All energies are measured in units of
$J_{\pm} = 1$.} \label{fig:fig3}
\end{center}
\end{figure*}

\subsubsection{Parton formulation of ordered states}
\label{sec:part-form-order} The PSG wave functions can be used as a
starting point on which magnetic order can be added. This is done by
adding a site dependent magnetic field $\vec{h}_i$ to the mean-field
Hamiltonians, which define our trial states:
\begin{eqnarray}
\mathcal{H}_{mo} =\mathcal{H}_{mf} \, - \,
\sum_i \vec{h}_i \cdot \vec{S}_i. \label{mfHam}
\end{eqnarray}
Magnetic order can be induced in this way on top of any of the 6 QSL
states.  In practice, the lowest energies are found by using
$\mathcal{H}_{mf}^{(1B)}$, i.e., by perturbing the Dirac spin
liquid. Notably, the Zeeman term in this case fully gaps the
partons. Consequently, the usual Polyakov argument, which applies to
an emergent $U(1)$ gauge theory with fully gapped Dirac fermions in
two dimensions, implies that monopole instantons proliferate and the
Dirac spinons are confined.  Thus, the projected wavefunction built
from $\mathcal{H}_{mo}$ describes a state adiabatically connected to
a conventional magnetically ordered one.

If $|\vec{h_i}|\rightarrow \infty$, Eq.~\eqref{mfHam} describes
classical magnetic order with $|\langle \vec{S}_i \rangle |=1/2$ on
each site.  If instead a finite field is used, the value of the
magnetic moment can be greatly reduced. In general, the energy
should be optimized with respect to the full {\em set} of Zeeman
fields $\vec{h}_i$ on all sites.  In practice, such an optimization
would have too many parameters. Instead, we guess an appropriate
pattern for these fields, and then optimize $|h|/t$ to give the
lowest variational energy.   For example, in the Heisenberg limit,
we choose the field to have a constant magnitude but an orientation
with a three-sublattice structure of total vector sum zero (the
symmetry pattern of the $120^\circ$ state):
\begin{eqnarray}
\vec{h}_i = |h|( \cos(\vec{q}\cdot \vec{r}_i+\phi),\sin(\vec{q}\cdot\vec{r}_i+\phi),0),\label{zeemanform}
\end{eqnarray}
where $\vec{q}$, $|h|$ and a phase $\phi$ are variational
parameters. In this case, the optimal magnetic field of our simple
ansatz gives a staggered magnetic moment $|\langle \vec{S}_i
\rangle| \approx 0.30$, while the corresponding DMRG calculations
for the triangular-lattice Heisenberg model find a staggered
magnetic moment $M\sim0.20$ \cite{PhysRevB.92.041105}. Including
local correlations in our variational state, for example, by
including Jastrow factors, will in general reduce the value of
$\langle S \rangle$ further. It is interesting that our PSG analysis
provides a general way to construct any ansatz satisfying the
constraint of Eq.~\eqref{constraint}, even allowing us to construct
energetically competitive \emph{magnetic} states in addition to
giving a general classification of all spin liquid states.

\subsubsection{Extended model}
\label{sec:extended-model}

\begin{figure}[t]
\begin{center}
{\bf a)} \includegraphics[scale=0.20]{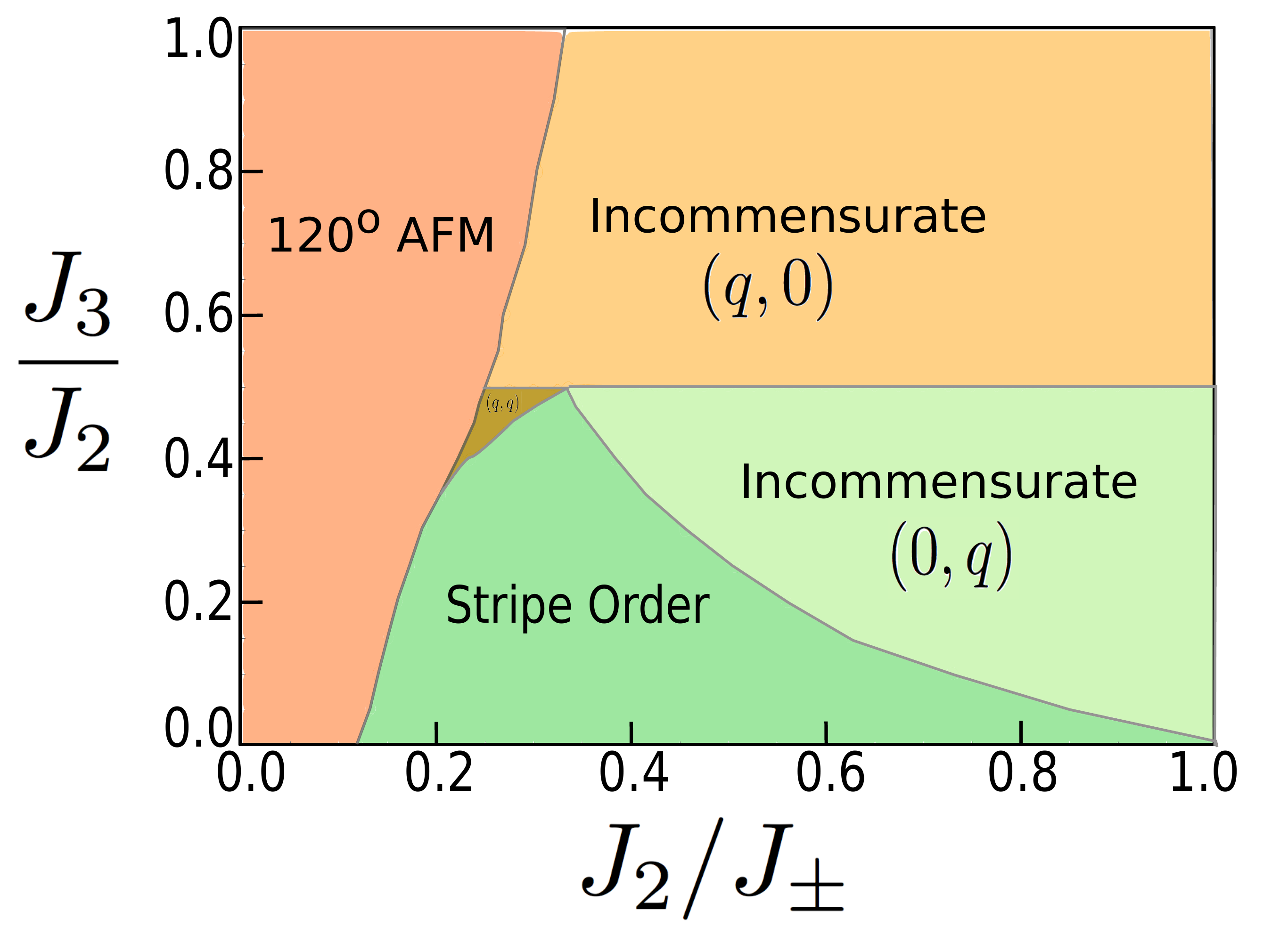}
\hspace{10mm} {\bf b)}\includegraphics[scale=0.206]{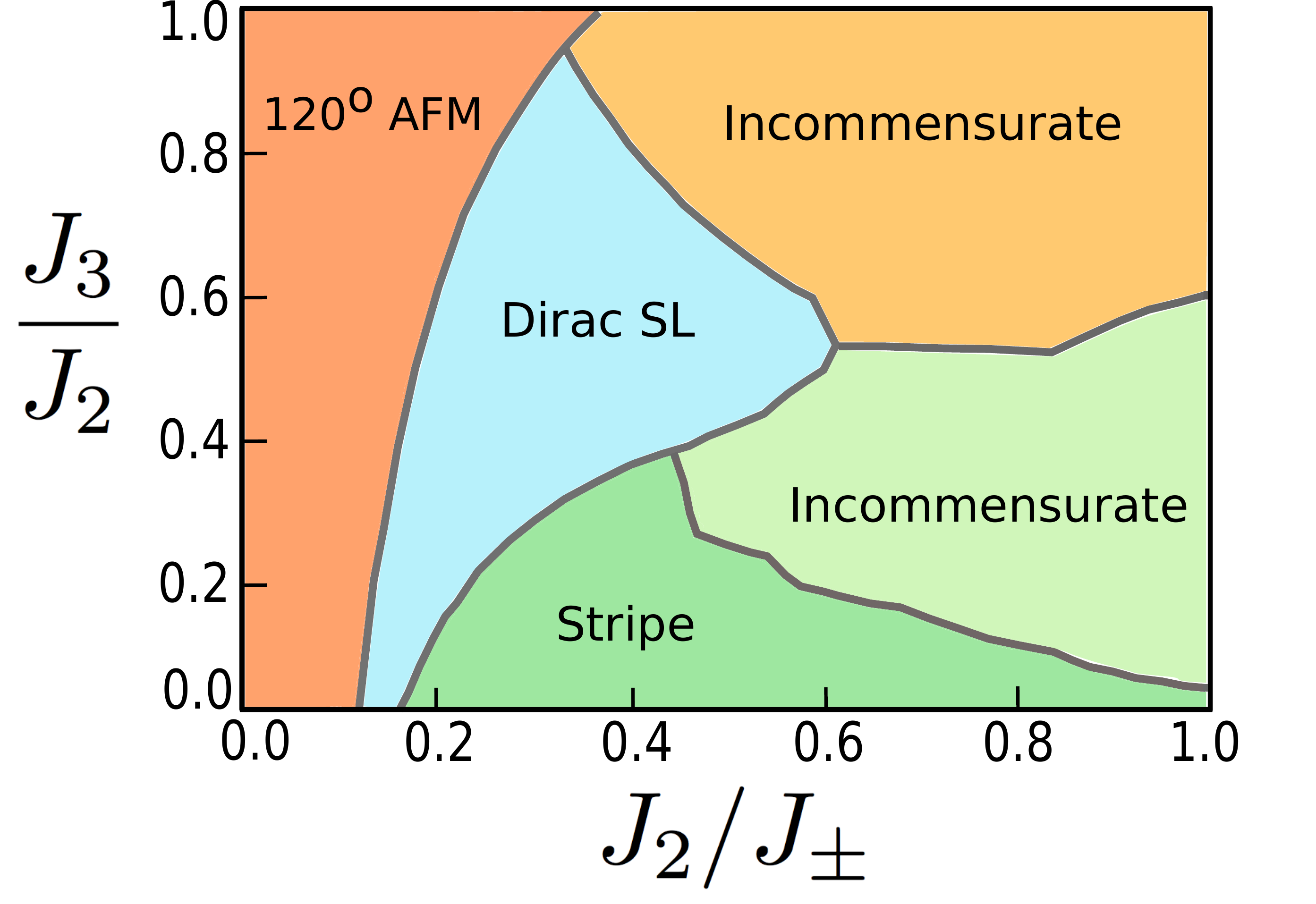}
\caption{{\bf a)} The classical phase diagram from the
Luttinger-Tisza method and {\bf b)} the same quantum phase diagram
from variational Monte Carlo at $J_z / J_\pm = 1$ and $J_{\pm \pm} =
J_{\pm z} = 0$.} \label{fig:fig4}
\end{center}
\end{figure}

Implementation of the above method shows that the nearest-neighbor
Hamiltonian $H_{nn}$ in Eq.~(\ref{Ham}) is dominated by magnetic
order. To find actual spin liquid physics, we therefore extend the
model to include second- and third-neighbor interactions. Keeping
the same relative XY anisotropy, we study the Hamiltonian:
\begin{eqnarray}
H &=& H_{nn} \, + \,  J_2 \sum_{\langle \langle ij \rangle \rangle} \left (S_i^+ S_j^- + S_i^- S_j^+ \, + \,
                    \frac{J_z}{J_\pm} S_i^z S_j^z \right ) \nonumber\\
        && \,+ \, J_3 \sum_{\langle \langle \langle ij \rangle \rangle \rangle } \left (S_i^+ S_j^- + S_i^- S_j^+ \, + \,
                    \frac{J_z}{J_\pm} S_i^z S_j^z \right ).
\label{Ham2}
\end{eqnarray}
To avoid complications involving canted magnetic orders, we restrict
our attention to the case of $J_{\pm z} = 0$.  With this in mind, in
this section, we undertake the somewhat ambitious goal of describing
the entire four-dimensional phase diagram in terms of the free
parameters $J_z$, $J_{\pm\pm}$, $J_2$, and $J_3$, all relative to
$J_\pm = 1$.

We first review what is already known about the ground state phase diagram of Eq.~(\ref{Ham2}):
\begin{itemize}
\item In the absence of second- and third- neighbor interactions ($J_2 = J_3 = 0$), the Luttinger-Tisza analysis of Ref.~\cite{PhysRevB.94.035107} tells us the magnetic order when $\vec{S}$ is treated as a classical vector. In that case, there is a phase transition from the 120$^\circ$ staggered antiferromagnetic state [ordered at wavevector $\vec{q}_{120} = (\frac{4\pi}{3},0)$] at small $|J_{\pm\pm}|$ to a striped phase [ordered at wavevector $\vec{q}_s = (0,\frac{2\pi}{\sqrt{3}})$] for $|J_{\pm\pm}| \gtrsim 0.25$.
\item There is also a great deal of literature on the quantum $J_1-J_2$ model ($J_{\pm\pm}=J_3=0$), in the Heisenberg limit
$(J_z = 2J_{\pm})$ \cite{PhysRevB.92.140403, PhysRevB.92.041105}. In this case, growing evidence suggests that a spin liquid phase interpolates between the
120$^\circ$ phase for small $J_2$ and the stripe phase at large $J_2$.
\end{itemize}

\subsubsection{VMC results}
\label{sec:variational-results}

\begin{figure*}[t]
\begin{center}
\includegraphics[scale=0.2455]{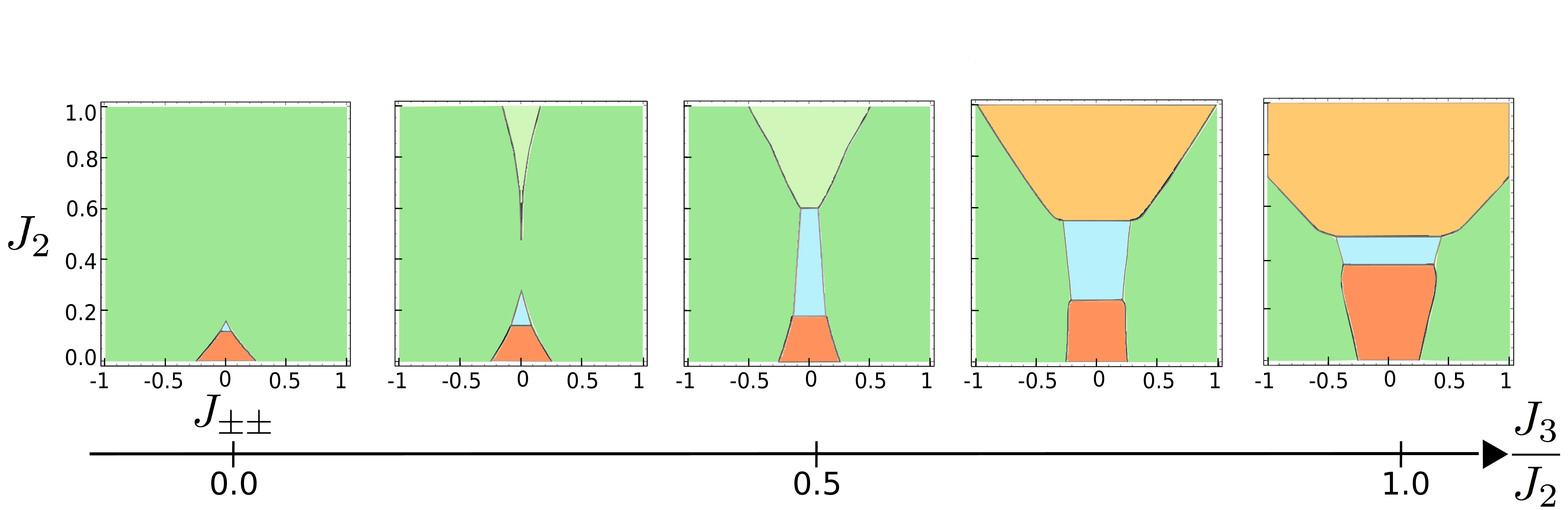}
\caption{The full $J_{\pm\pm}-J_2-J_3$ quantum phase diagram for
$J_z = 1$ and $J_{\pm z} = 0$. Note that the color scheme is the
same as in Fig.~\ref{fig:fig4}. Third neighbor interactions $J_3$
strongly disfavor stripe order (dark green) and increase the range
of the spin liquid phase (light blue). The horizontal axis on each
subplot gives the value of $J_{\pm\pm}$. All energies are measured
in units of $J_{\pm} = 1$.} \label{fig:fig5}
\end{center}
\end{figure*}

The advantage of using variational Monte Carlo with simple trial
wave functions is that we are able to explore a huge phase space of
our Hamiltonian. We consider several ans\"atze for magnetic order,
taking the Zeeman field in the form of Eq.~\eqref{zeemanform} with
wavevector $\vec{q}_v = (q,0)$ or $\vec{q}_v = (0,q)$, where $q$,
$|h|$, and a phase $\phi$ are variational parameters, which allows
for both commensurate and incommensurate ordering. In practice, we
find that the energies of all our ans\"atze, except for the striped
phase with $\vec{q}_s = (0, \frac{2\pi}{\sqrt{3}})$, are independent
of $\phi$, even when the $U(1)$ symmetry is broken by $H_{\pm\pm}$.
In the stripe phase, we find that the minimum energy is always
obtained for $\phi=0$ when $J_{\pm\pm}>0$, giving the ordering
pattern seen in Fig.~\ref{fig:fig1}b), and for $\phi=\pi/2$ when
$J_{\pm\pm}<0$, which rotates all spins by 90$^\circ$. In
Fig.~\ref{fig:fig3}, we present our result for the full quantum
$J_z-J_{\pm\pm}-J_2$ phase diagram. Notice that our results agree
well with the previously understood limits. When $J_2= 0$, the
system acts very similar to the classical case, with a transition
between the 120$^\circ$ and stripe orders around $J_{\pm\pm}\approx
0.20 + 0.05 J_z$. When a second-neighbor interaction is added, we
indeed see that a Dirac spin liquid appears between the 120$^\circ$
and stripe phases.  This phase is stable for small $J_{\pm\pm}$, but
both large $J_2$ and $J_{\pm\pm}$ favor stripe order, leading to the
triangular shape of the spin liquid regime which we see in
Fig.~\ref{fig:fig3}. It is also notable that the extent of the spin
liquid phase shrinks dramatically when $J_z$ is lowered from the
Heisenberg point.  This is in agreement with the DMRG results on
this model in Ref.~\cite{2017arXiv170302971Z}.

We are also able to go beyond this model to look at the effect of
the third-neighbor XXZ interaction $J_3$. Since both the second- and
third-neighbor sites are separated by two lattice bonds, a simple
superexchange picture implies that such a term would be present in
materials with $J_3 \sim 0.5 J_2$. We will see that the effect of
such a term is to enhance the size of the spin liquid regime.

First, we present the results in the classical limit. When
$J_{\pm\pm}=0$, the system has $U(1)$ symmetry and we can solve for
the classical magnetic order using the Luttinger-Tisza method since
any coplanar magnetic order with a single ordering wave vector will
satisfy the hard constraint that $\vec{S}_i = 1/2$ on every site.
The result is that, in addition to the usual $120^\circ$ and stripe
phases, $J_3$ favors two additional incommensurate magnetic phases,
with ordering wave vectors at $(q,0)$ and $(0,q)$.  These phases can
be thought of as the incommensurate versions of the 120$^\circ$ and
stripe phases, respectively. A third incommensurate order with wave
vector $(q,q)$ also appears classically, but we will ignore this as
such a phase never appears in the quantum case. The full classical
phase diagram is shown in Fig.~\ref{fig:fig4}a) and is independent
of $J_z$ since the ordering is always in the $xy$ plane.

Our VMC results on the quantum model agree remarkably well with the
classical phase diagram, considering we have used completely
different methods. Fig.~\ref{fig:fig4}b) shows the results for $J_z
= 1.0$.  We see that the shapes of the magnetic phases are largely
the same as in the classical case, but the intermediate region where
the phases meet is occupied by a broad spin liquid regime.

In Fig.~\ref{fig:fig5}, we show the full $J_2-J_3-J_{\pm\pm}$ phase
diagram for $J_z = 1.0$. In addition to the presence of
incommensurate magnetic order, the major feature of the data is that
the spin liquid regime is enhanced with respect to the $J_3=0$ case.
The third-neighbor interaction provides further frustration and
finds stripe order particularly unfavorable. The spin liquid phase
therefore survives to a large value of $J_{\pm\pm}$ when $J_3$ is
included.

As mentioned previously, more accurate energies can be found by
adding further variational parameters to the wave function, such as
allowing for Jastrow factors \cite{PhysRevB.71.241103,
PhysRevB.64.024512} or performing a small number of L\'anczos steps
\cite{lancos1}. However, we find that supplementing the PSG wave
functions in this way only gives small improvements in the energies,
leading to very small shifts of the phase boundaries. In section IV,
we look at how we can make {\it qualitative} changes to the spin
liquid ans\"atze.

In summary, our variational Monte Carlo calculation allowed us to
explore a huge parameter space of the Hamiltonian in
Eq.~(\ref{Ham2}) and to obtain quantitative results for the ground
state in each parameter regime. When a second-neighbor interaction
is added, the \emph{Dirac} spin liquid appears as the ground state
between the 120$^\circ$ and stripe phases. This phase shrinks
dramatically away from the Heisenberg limit, but is in fact enhanced
when a small third-neighbor interaction is included.

\section{Beyond the PSG Wave Function}

\subsection{Perturbative Correction to the Wave function}

In this section, we take a more phenomenological approach to
studying a quantum spin liquid in the presence of strong spin-orbit
coupling. We propose modifications to the mean-field ans\"atze which
can be implemented numerically in the variational wave functions.

The plain mean field ans\"atze are limited in the amount of
complexity they can accommodate. The main issue with the VMC
simulation in this context is that the two most energetically
competitive states, the Fermi surface and the Dirac spin liquid
ones, possess \emph{too} much symmetry. Our trial wave functions
have no coupling between the spin and orbital degrees of freedom,
which is a feature one would expect to find in the Hamiltonian's
true ground state. Furthermore, according to the PSG analysis, no
fermion bilinear operators inducing such spin orbit coupling can be
added to the uniform Fermi surface Hamiltonian, not even at the
further-neighbor level.

Instead, we formulate a method to incorporate {\it many-body}
effects which modify our wave functions. Inspired by the path
integral formulation for an interacting quantum field theory, we
consider the variational state
\begin{eqnarray}
\ket{\Psi} = e^{-\alpha {\sf H}} \, \mathcal{P} \ket{\psi_0}
\label{WF},
\end{eqnarray}
where ${\sf H} = {\sf H}_{\pm \pm}$ is defined in Eq.~(\ref{Ham}).
This form is reminiscent of the L\'anczos algorithm, where
applications of large powers of an operator project a trial state
into the ground state of the given operator. Indeed, if we let
$\alpha \rightarrow \infty$, this operator projects into the ground
state of ${\sf H}$. Instead, however, we take a slightly different
approach, and let $\alpha$ be a small perturbation on $\mathcal{P}
\ket{\psi_0}$, treating $\ket{\Psi}$ as a variational wave function.

There have been previous works combining the L\'anczos algorithm
with variational Monte Carlo \cite{PhysRevB.93.144411,lancos1}. This
proceeds by applying a finite number of L\'anczos steps and working
with the wave function $\ket{\Psi^{(n)}} = (1 + \sum_{p=1}^n
\alpha_p H^p) \ket{\psi_0}$, where the series is truncated for some
small $n$, and the $\alpha_p$ are left as variational parameters.
While this works well if the initial state is very close to the
ground state of $H$, it is less effective as a phenomenological
tool. The reason is that corrections at any finite order $n$
necessarily scale to zero in the thermodynamic limit. When
calculating the \emph{correction} to an expectation value using
$\ket{\Psi^{(n)}}$, ``disconnected'' powers of the Hamiltonian are
subtracted off in the numerator, but not in the denominator. The
normalization factor in the denominator therefore necessarily grows
faster than the numerator with system size. Additional powers of $n$
are then needed to compensate for this fact, but a fully extensive
correction is only found at $n\sim N$.

Instead, we have found that the best way to work with the wave
function in Eq.~(\ref{WF}) numerically is to implement the
correction perturbatively in $\alpha$, but to all powers in $n$. To
do this, we realize that the expectation value of any operator with
respect to our improved wave function can be written as
\begin{eqnarray}
\left \langle \mathcal{O} \right \rangle = \frac{ \left \langle
e^{-\alpha {\sf H}} \mathcal{O} e^{-\alpha {\sf H}} \right
\rangle_0}{ \left \langle e^{-2\alpha {\sf H}} \right \rangle_0}
\label{Op-1},
\end{eqnarray}
where $\langle \cdots \rangle_0$ is the expectation value taken with
respect to the unperturbed wave function $\mathcal{P}\ket{\psi_0}$.

It is now possible to expand Eq.~(\ref{Op-1}) analogously to
diagrammatic perturbation theory. For any Hermitian operator
$\mathcal{O}$, the expanded correction reads
\begin{align}
\left \langle \mathcal{O} \right \rangle = \frac{ \left ( \langle \mathcal{O}\rangle_0 - 2
\alpha \, \mathrm{Re} [\langle \mathcal{O} \sf{H} \rangle_0] +
\alpha^2 \left( \langle \sf{H} \mathcal{O} H \rangle_0 + \mathrm{Re}
[\langle \sf{H}^2 \mathcal{O} \rangle_0 ] \right)  + O(\alpha^3) \right) } { \left
( 1 - 2 \alpha \langle \sf{H} \rangle_0 +  2 \alpha^2 \langle
\sf{H}^2 \rangle_0  + O(\alpha^3) \right )}. \label{Op-2}
\end{align}
The subtle difference is that now, by including all powers of $n$, all terms in the denominator exactly cancel the higher order
``disconnected" pieces in the numerator. In the VMC calculation,
this is expressed by the fact that $\langle H_{ij} H_{k\ell} \rangle
\approx \langle H_{ij} \rangle \langle H_{k\ell} \rangle$ as $|(ij)
- (kl)| \rightarrow \infty$. This way, we are able to measure, in
our numerical simulation, many-body corrections to the wave function
which survive in the thermodynamic limit.

In principle, applying the operator $\exp [-\alpha {\sf H}]$ to our
unperturbed trial wave function could cause a phase transition, and
we would no longer be working with a spin liquid state. For small
$\alpha$, however, we expect that the spin liquid ground state
should be stable to such a perturbation. In the spinon metal, in a
similar vein to Fermi liquid theory, we expect that these terms only
give a correction to the self-energy of spinons near the Fermi
surface \cite{joe1}.

\subsection{Correction to the Energy}

\begin{figure}[t]
\begin{center}
\hspace{-3mm} {\bf a)} \includegraphics[scale=0.10]{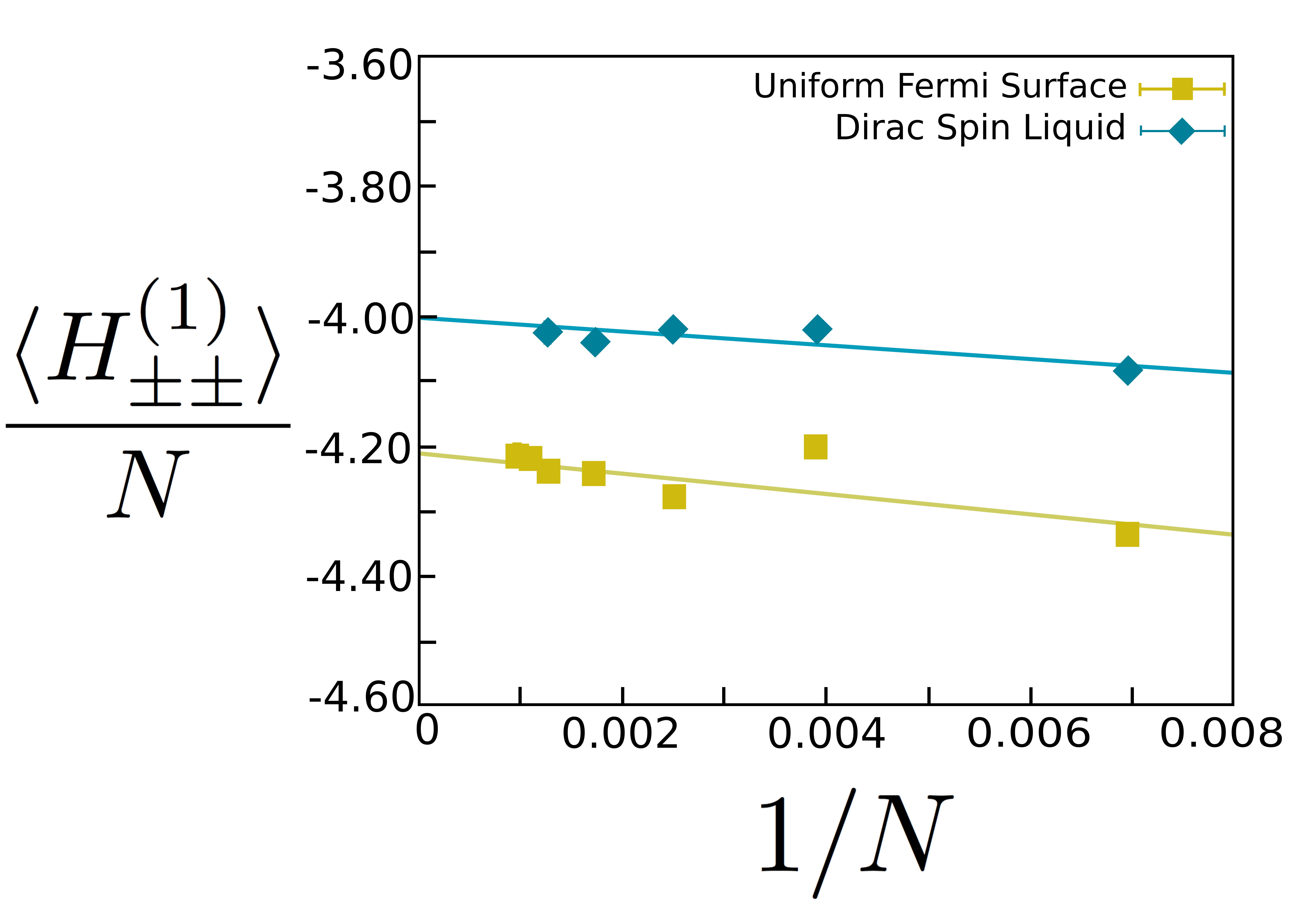}
\hspace{7mm} {\bf b)}  \includegraphics[scale=0.10]{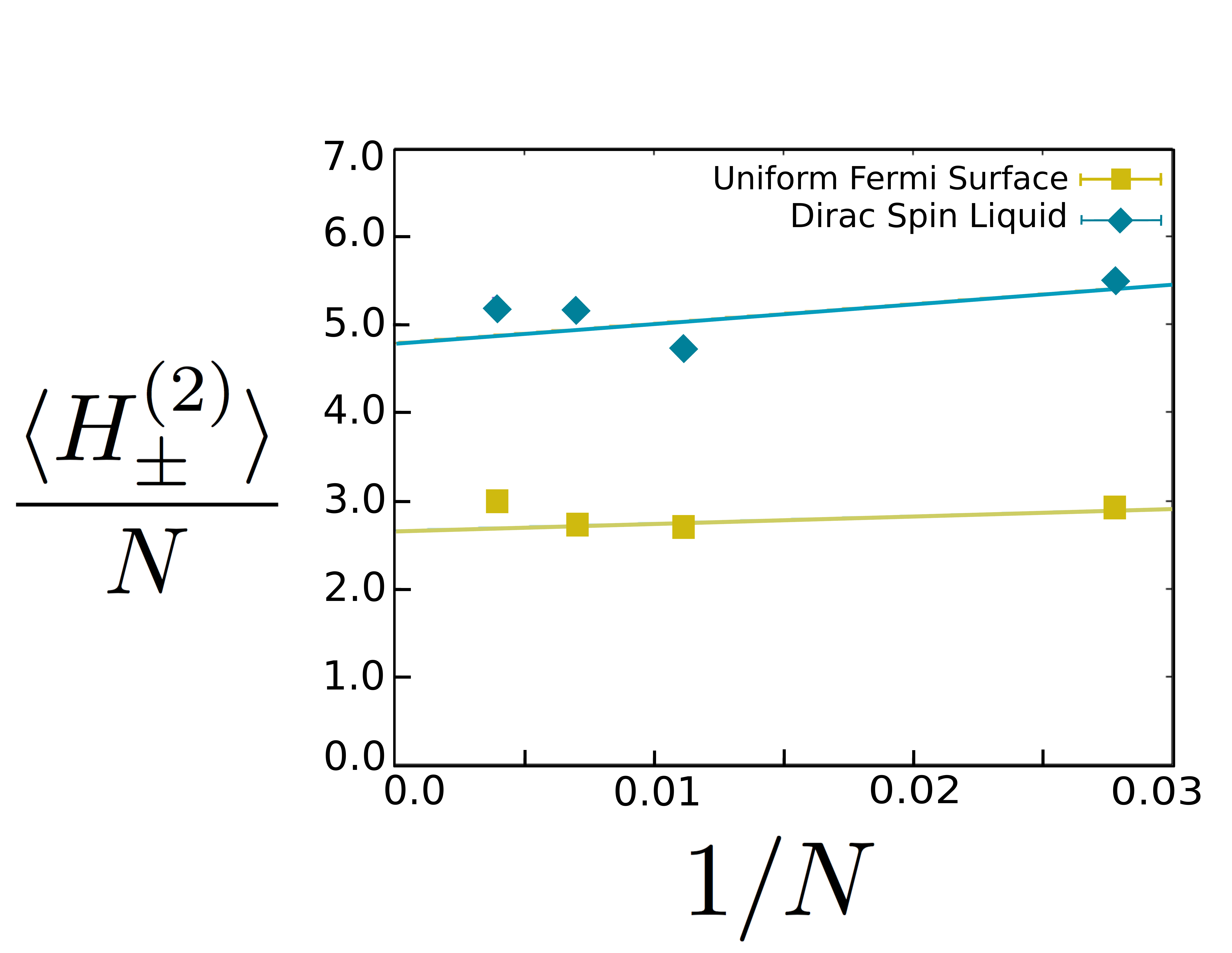}
\caption{Finite size scaling of the lowest-order correction to {\bf
a)} $\langle {\sf H}_{\pm\pm}\rangle$ and {\bf b)} $\langle {\sf
H}_{\pm} \rangle$, for both the uFS (yellow) and Dirac (blue) spin
liquid states. The corresponding change in energy is $\Delta E \sim
\alpha J_{\pm\pm} \langle {\sf H}_{\pm\pm} \rangle + \alpha^2
J_{\pm} \langle {\sf H}_{\pm} \rangle$. } \label{fig:fig6}
\end{center}
\end{figure}

To begin, we measure the correction to the energy of the Dirac and
uniform Fermi surface states, which arises from including the
spin-orbit interaction in our variational wave function.  We can
directly measure the first and second order corrections numerically.

For any operator $\mathcal{O}$, we write the n$^{th}$ order
correction to the expectation value $\langle \mathcal{O} \rangle$
from applying $\exp [-\alpha {\sf H}]$ as $\alpha^n \langle
\mathcal{O}^{(n)}\rangle$. Expanding Eq.~(\ref{Op-2}) gives
\begin{eqnarray}
\langle \mathcal{O}^{(1)} \rangle &=& -2 (\mathrm{Re} [ \langle
\sf{H} \mathcal{O} \rangle_0 ] -
\langle \sf{H} \rangle_0 \langle \mathcal{O} \rangle_0 ), \\
\langle \mathcal{O}^{(2)} \rangle &=& \langle \sf{H} \mathcal{O}
\sf{H} \rangle_0 + \mathrm{Re} [\langle \sf{H}^2 \mathcal{O}
\rangle_0 ]
- 4 \mathrm{Re} [\langle \sf{H} \rangle_0 \langle \mathcal{O} \sf{H} \rangle_0 ]\nonumber \\
&& - 2 \langle \sf{H}^2 \rangle_0  \langle \mathcal{O} \rangle_0 + 4
\langle \sf{H} \rangle_0^2 \langle \mathcal{O} \rangle_0. \nonumber
\end{eqnarray}
In our case, ${\sf H} = {\sf H}_{\pm\pm}$ and $\langle {\sf H}_{\pm
\pm} \rangle_0 = \langle {\sf H}_{\pm\pm} {\sf H}_{\pm} \rangle_0 =
0$. Therefore, the spin-orbit part of the Hamiltonian is altered at
order $\alpha$, while the rotationally invariant part is corrected
at order $\alpha^2$:
\begin{eqnarray}
\langle {\sf H}_{\pm\pm}^{(1)} \rangle  &=& -2 \langle {\sf H}_{\pm\pm}^2 \rangle_0 , \\
\langle {\sf H}_{\pm}^{(2)} \rangle  &=& \mathrm{Re} [\big \langle
\big \{ {\sf H}_{\pm\pm}, {\sf H}_{\pm} \big \} {\sf H}_{\pm\pm}
\big
\rangle_0 ] - 2 \langle {\sf H}_{\pm\pm}^2 \rangle_0 \langle {\sf H}_{\pm}\rangle_0, \nonumber \hspace{3mm} \\
\langle {\sf H}_z^{(2)}\rangle  &=& \mathrm{Re} [\big \langle \big
\{ {\sf H}_{\pm\pm}, {\sf H}_{z} \big \} {\sf H}_{\pm\pm}  \big
\rangle_0 ] - 2 \langle {\sf H}_{\pm\pm}^2 \rangle_0 \langle {\sf
H}_{z} \rangle_0. \nonumber
\end{eqnarray}
In Fig.~\ref{fig:fig6}, we show the resulting scaling of $\langle
{\sf H}_{\pm\pm}^{(1)} \rangle$ and $\langle {\sf
H}_{\pm}^{(2)}\rangle$ to the thermodynamic limit. The result is
that the spinon metal is more susceptible, compared to the Dirac
state, to energetically beneficial corrections to ${\sf H}_{\pm\pm}$
and less susceptible to detrimental corrections to ${\sf H}_{\pm}$
and ${\sf H}_{z}$. Putting this together, we find that the optimal
value of the variational parameter is $\alpha_{min} \sim J_{\pm\pm}
/ (J_{\pm}+J_z)$, which gives an energy correction $\Delta E \sim
-J_{\pm\pm}^2 / (J_{\pm}+J_z)$. More precisely, we find that the
energy densities after the lowest-order corrections are given by
\begin{eqnarray}
\text{E}_{uFS} / N = -0.4682(1+J_z/4) - \frac{1.56 \, J_{\pm\pm}^2}{J_{\pm} + 1.42 J_z}, \nonumber \\
\text{E}_{Dirac} / N = -0.7050(1+J_z/4)- \frac{0.84 \,
J_{\pm\pm}^2}{J_{\pm} + 0.87 J_z}.
\end{eqnarray}
This implies that that the Fermi surface state becomes energetically
superior to the Dirac state between $J_{\pm\pm}=0.57$ at $J_z=0$ and
$J_{\pm\pm} = 1.54$ at $J_z=2.0$. One caveat, of course, is that
these values of $J_{\pm\pm}$ may fall outside the perturbative
regime. Also, while smaller $J_z$ appears to be more favorable for
the spinon Fermi surface, this is also the parameter regime which is
more susceptible to magnetic order.

\subsection{Correction to the Spin Structure Factor}

\begin{figure}[t]
\begin{center}
\includegraphics[scale=0.17]{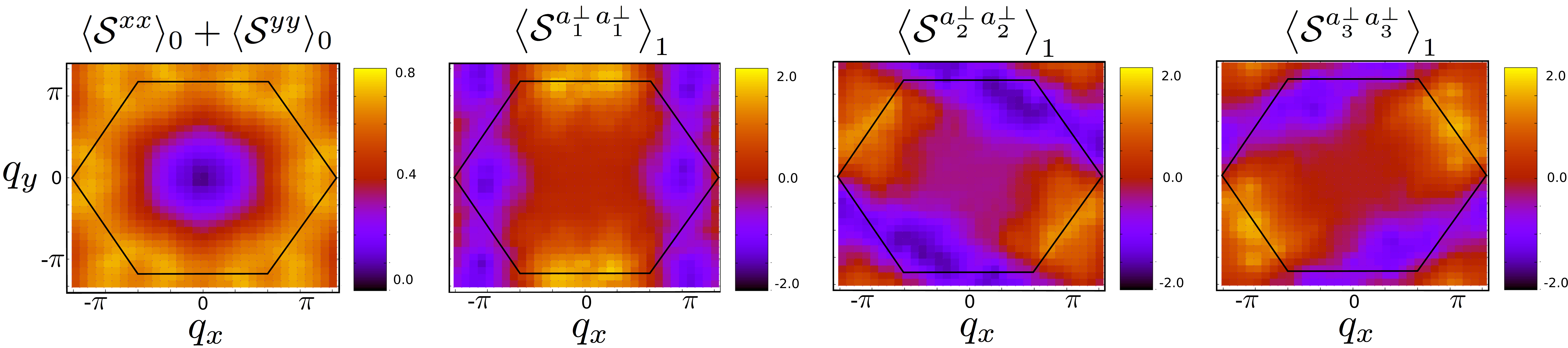}
\caption{The rotationally invariant spin structure factor (top left)
and the perturbative corrections to the spin-polarized structure
factors measured with spins pointing perpendicular to the three
lattice bond directions $\vec{a}_1,\vec{a}_2$, and $\vec{a}_3$,
within the plane of the triangular lattice.} \label{fig:fig7}
\end{center}
\end{figure}

Studying the improved variational wave function makes it clear that
the spinon metal state in a spin-orbit coupled environment has
several unique properties, despite the fact that the mean-field
Hamiltonian retains its rotational invariance.  Taking our analogy
to Fermi liquid theory seriously, the spin-orbit interaction gives a
momentum and spin dependent correction to the self energy. This
appears as a momentum dependent correction to the structure factor,
which we can again measure directly in our simulation.

We differentiate between the various spin polarized contributions to
the spin-spin correlation function:
\begin{eqnarray}
\mathcal{S}^{\alpha \beta}(\vec{q}) = \sum_{i} e^{i \vec{q} \cdot
\vec{r}_i}\langle S^\alpha_i S^\beta_0 \rangle. \label{Sab}
\end{eqnarray}
The first-order correction to the correlation function is
\begin{equation}
\langle S_i^\alpha S_j^\beta \rangle_1 = -2 \left[
\mathrm{Re}[\langle S_i^\alpha S_j^\beta {\sf H}_{\pm\pm} \rangle_0]
- \langle S_i^\alpha S_j^\beta \rangle_0 \langle {\sf
H}_{\pm\pm}^{\phantom{\beta}} \rangle_0 \right].
\end{equation}
The results are shown in Fig.~\ref{fig:fig7}. The corrections to the
spin-polarized structure factor are direction-dependent broad peaks
at the $M$ points of the Brillouin zone which appear at order
$\alpha \sim J_{\pm\pm}/(J_{\pm}+J_z)$. Therefore, in a spinon metal
with spin-orbit coupling, spin-spin correlations when measured with
different spin polarizations are direction dependent. This type of
measurement could prove to be an important test to show both the
presence of spin-orbit interactions and the absence of spontaneous
symmetry breaking. Similar directional peaks can be seen in related
models when spin-orbit terms are directly included in the ground state
ansatz \cite{PhysRevB.95.024421}.
We note that these kinds of direction-dependent
structure factors have already been measured experimentally by
resonant elastic x-ray scattering in the honeycomb lattice iridate
Na$_2$IrO$_3$ \cite{bjkim}.

\subsection{Thermal Hall Conductivity}

\subsubsection{General considerations}
\label{sec:generalities}

Thermal transport measurements can be a powerful tool for studying
magnetic insulators. The idea is to set up a thermal gradient
$\nabla T$ (which is analogous to an electric field) and then
measure the heat current $j^{\text{th}}$ in response to it (which is
analogous to an electric current). Any heat current in the insulator
must be carried by the emergent quasiparticles, giving us a probe of
the low energy excitations. The thermal conductivity, $\kappa$, can be
defined within linear response as
\begin{eqnarray}
j_\mu^{\text{th}} = -\kappa_{\mu \nu} \partial_\nu T.
\end{eqnarray}
The spinon Fermi surface QSL is unusual due to the large number of
gapless excitations. This leads to a predicted linear $T$ term
appearing in the  diagonal component of $\kappa$, similar to what
one would expect in a metal.  The deconfined spinons carry heat in
the same way physical electrons carry charge in an electrical
conductor. A major difficulty is that many degrees of freedom, most
notably phonons, can contribute to the diagonal thermal
conductivity, making the measurement challenging.

The thermal \emph{Hall} conductivity, however, given by the
off-diagonal component of $\kappa$, should not contain a phonon
term. Furthermore, as explained in
Ref.~\cite{PhysRevLett.104.066403}, it is very difficult to find an
effect generated by magnons on the triangular lattice due to a
cancellation of the contributions from neighboring edge sharing
plaquettes.  A large nonzero thermal Hall conductivity could
therefore be a strong indicator of exotic physics. Indeed, in
Ref.~\cite{PhysRevLett.104.066403}, the authors also predict that a
spinon metal would display such an effect. However, the reasoning is
very subtle, depending on a coupling of the \emph{orbital} motion of
the spinons to the external electromagnetic field through the
interaction with the internal gauge field.

Here, we argue that there exists a distinct contribution to the
thermal Hall conductivity in the spinon metal which is unique to
spin-orbit coupled systems and relies only on a Zeeman coupling to
the external electromagnetic field.  For itinerant fermions with
conserved charge,  the presence of spin-orbit coupling can lead to a
nontrivial Berry curvature which may induce an anomalous component
of the charge Hall conductivity, in the absence of any Lorentz
force. This mechanism of anomalous Hall conductivity was explored
intensely for Rashba two-dimensional electron gases and in many
other models. In the following, we adapt this idea to study the {\em
thermal} conductivity of the Fermi surface QSL state.

The U(1) QSL states studied here have an {\em emergent} conserved
charge, which is the fermion number associated with the emergent
U(1) gauge symmetry.  Consequently, at the parton level, we can
define a current associated with this charge, and we may consider,
formally, the emergent conductivity tensor $\sigma^{qp}_{\mu\nu}$
defined with respect to the emergent current and a potential
coupling to the associated emergent charge density.  This is not the
true electrical conductivity, since this is an insulator, and it is
also not the thermal conductivity.  Thus we proceed in two stages.
First, we consider the anomalous emergent Hall conductivity of the
spinons.  Then, we relate it to the more easily measurable {\em
thermal} Hall conductivity (in principle, the emergent conductivity
should also be measureable, but it is not obvious how to do so).

\subsubsection{Effective quasiparticle Hamiltonian}
\label{sec:effective-hamiltonian}

At the mean field level, the emergent Hall conductivity can be
extracted as an integral over the Berry curvature of the occupied
spinon bands.  Within the simple PSG wave function, the spinon metal
is spin-rotationally invariant and therefore has zero Berry
curvature. On symmetry grounds, however, we expect that a Hall
conductivity should microscopically arise.  To estimate it, we
consider the `improved' wave function, and infer a self-energy
correction which breaks spin-rotational symmetry and induces a
non-zero Berry curvature.

The Berry gauge field (Berry connection) is defined for a single
particle system as
\begin{eqnarray}
\vec{A} (k) &=& -\im \,  \langle u_k | \vec{\nabla}_k |u_k \rangle,
\end{eqnarray}
where $\ket{u_k}$ is defined as in the Bloch wave function. The
anomalous Hall conductivity is then given by
\begin{eqnarray}
\sigma^{qp}_{xy} &=& \oint_{\partial S} \vec{A} (k) \cdot d\vec{k}
\, = \, \int_S \, [\vec{\nabla}_k \times \vec{A} (k)] \, d^2k,
\end{eqnarray}
where the first (line) integral is taken around the Fermi surface
$\partial S$, while the second (area) integral is taken over the
area $S$ spanned by it. This physical quantity is invariant under
$U(1)$ gauge transformations, as is immediately evident from its
expression in terms of the Berry curvature $\mathcal{B} (k) =
\vec\nabla_k \times \vec{A} (k)$.

To obtain the Berry curvature, we suppose that the system is
described by an effective quasiparticle Hamiltonian including a
self-energy correction $\Sigma(k)$ and a Zeeman coupling to an
external magnetic field $\vec{B} = h \hat{z}$:
\begin{eqnarray}
\mathcal{H}_{\text{eff}} (k) \, = \, \left( f^\dagger_{k \uparrow}
\, f^\dagger_{k \downarrow} \right) \left( \begin{array}{cc}
\varepsilon (k) - h & \Sigma^* (k) \\
\Sigma (k)  & \varepsilon (k) + h
\end{array} \right)
\left( \begin{array}{c} \f_{k \uparrow} \\ \f_{k \downarrow}
\end{array} \right). \nonumber \\ \label{eq:2}
\end{eqnarray}
We determine the self-energy $\Sigma(k)$ by requiring that the
off-diagonal expectation value   $\Pi_{\uparrow \downarrow}(k)
\equiv \langle f^\dagger_{k \uparrow} f_{k \downarrow}\rangle$
calculated using the improved wave function {\em matches} that
calculated using the effective Hamiltonian $\mathcal{H}_{\text{eff}}
(k)$.

To proceed, we consider an improved wave function similar to that in
Eq.~\eqref{WF}:
\begin{equation}
\label{eq:1} \ket{\Psi} = e^{-\tilde\alpha \tilde{\sf H}} \,
\mathcal{P} \ket{\psi_0},
\end{equation}
where now we take $\tilde{\sf H} = {\sf H}_{\pm z}$.  The reason for
this change is that the previously-considered correction due to
${\sf H}_{\pm\pm}$ gives exactly zero contribution to $\Pi_{\uparrow
\downarrow}$ because it conserves the total spin $S^z$ modulo 2.
The analogous contribution due to ${\sf H}_{\pm z}$, however, does
contribute.  We expect that the energetically optimal value of the
variational parameter is $\tilde\alpha \sim J_{\pm z} / J_0$, where
$J_0$ is on the order of the other exchange couplings ($J_{\pm}$ and
$J_z)$.

Using the same perturbative expansion as above, the first-order form
of $\Pi_{\uparrow \downarrow} (k)$ becomes
\begin{eqnarray}
\Pi_{\uparrow \downarrow} (k) &=& \left \langle e^{-\tilde\alpha
\tilde{\sf H}} f_{k \uparrow}^\dagger \f_{k \downarrow}
e^{-\tilde\alpha \tilde{\sf H}}
\right\rangle_0 \nonumber \\
&&\hspace{-15mm} = -\tilde\alpha \Big ( \langle f_{k
\uparrow}^\dagger \f_{k \downarrow} \tilde{\sf H} \rangle_0 +
\langle \tilde{\sf H} f^\dagger_{k \uparrow} \f_{k \downarrow}
\rangle_0 \Big ) \nonumber \\
&& \hspace{-15mm} \equiv \Pi^{(1)}_R (k) + \Pi^{(1)}_L (k).
\label{Hn0}
\end{eqnarray}
If we represent the spin-spin interaction in momentum space with a
momentum-dependent form factor
\begin{eqnarray}
\tilde{\gamma} (k) \,=\, \frac{i}{2} \sum_{\mu = 1}^3 \sum_{\pm}
\gamma_\mu e^{\pm i \vec{k} \cdot \vec{a}_\mu}, \qquad \gamma_\mu
\equiv \gamma_{\vec{0}, \vec{a}_\mu},
\end{eqnarray}
the first expectation value $\Pi^{(1)}_R (k)$ takes the form
\begin{eqnarray}
\Pi_R^{(1)}(k) &=& i \tilde\alpha \Big \langle f_{k
\uparrow}^\dagger \f_{k \downarrow} \sum_{\langle m n \rangle}
\left[ \gamma_{mn}^{\phantom{z}} S^z_m S^-_n + (m \leftrightarrow n) \right] \Big \rangle_0 \nonumber \\
&& \hspace{-10mm} = \frac{\tilde\alpha} {N} \sum_{k_1, k_2, k_3}
\Big[ \langle f^\dagger_{k \uparrow} \f_{k \downarrow}
f^\dagger_{k_1 \uparrow} \f_{k_2 \uparrow} f^\dagger_{k_3 \downarrow} \f_{(k_1 - k_2 + k_3) \uparrow} \rangle_0 \nonumber \\
&& \hspace{-6mm} - \, \langle f^\dagger_{k \uparrow} \f_{k
\downarrow} f^\dagger_{k_1 \downarrow} \f_{k_2 \downarrow}
f^\dagger_{k_3 \downarrow} \f_{(k_1 - k_2 + k_3) \uparrow} \rangle_0
\Big] \, \tilde{\gamma} (k_1 - k_2) \nonumber \\
&& \hspace{-10mm} = -\frac{\tilde\alpha} {N} \sum_q \Big[ \langle
f^\dagger_{k \uparrow} \f_{k \uparrow} f^\dagger_{q \uparrow}
\f_{q \uparrow} \rangle_0 \, \langle \f_{k \downarrow} f^\dagger_{k \downarrow} \rangle_0 \nonumber \\
&& \hspace{-6mm} + \, \langle f^\dagger_{k \uparrow} \f_{k \uparrow}
\rangle_0 \, \langle \f_{k \downarrow} f^\dagger_{k \downarrow}
\f_{q \downarrow} f^\dagger_{q \downarrow}\rangle_0 \Big] \,
\tilde{\gamma} (k - q),
\end{eqnarray}
where we arrive at the last line after conserving spin and momentum
in the zeroth-order expectation values as well as using
$\tilde{\gamma} (-k) = \tilde{\gamma} (k)$ and $\tilde{\gamma} (0) =
0$.

Performing similar manipulations on $\Pi_L^{(1)} (k)$ and combining
the two contributions gives
\begin{eqnarray}
\Pi_{\uparrow \downarrow} (k) &=& \Pi_R^{(1)} (k) + \Pi_L^{(1)} (k)
= -\tilde\alpha \Lambda \tilde\gamma (k) \Gamma (k),
\nonumber \\ \nonumber \\
\Gamma (k) &=& \langle n_{k \uparrow} \rangle_0 \langle 1 - n_{k
\downarrow} \rangle_0 + \langle n_{k \downarrow} \rangle_0
\langle 1 - n_{k \uparrow} \rangle_0 \nonumber \\
&=& \coth (h / T) \left[ \langle n_{k \uparrow} \rangle_0 - \langle
n_{k \downarrow} \rangle_0 \right],
\label{H0} \\ \nonumber \\
\Lambda &=& \frac{1}{N} \sum_q e^{\pm i \vec{q} \cdot \vec{a}_\mu}
\left[ \langle n_{q \uparrow} \rangle_0 + \langle 1 - n_{q
\downarrow} \rangle_0 \right] \nonumber \\
&\sim& a^2 \int d^2 q \, e^{\pm i \vec{q} \cdot \vec{a}_\mu} \left[
\langle n_{q \uparrow} \rangle_0 - \langle n_{q \downarrow}
\rangle_0 \right], \nonumber
\end{eqnarray}
where $n_{k \sigma} = f^\dagger_{k \sigma} \f_{k \sigma}$ is a
number operator and $a = |\vec{a}_\mu|$ is the lattice constant.
Importantly, $\Lambda$ is real and independent of both $\mu$ and
$\pm$ due to the sixfold symmetry $\mathcal{S}_6$. Furthermore, in
the limit of $T \ll |h|$, the integrand is only non-zero in an
annulus of thickness $\sim h / (a J_0)$ around the Fermi surface of
radius $\sim 1/a$, and the integral can then be estimated as
$\Lambda \sim h / J_0$.

Let us also calculate $\Pi_{\uparrow \downarrow}(k)$ from the
effective Hamiltonian in Eq.~\eqref{eq:2}. In the limit of $| \Sigma
(k) | \ll |h|$, we obtain
\begin{eqnarray}
\Pi_{\uparrow \downarrow}(k) &=& -\frac{\mathrm{sgn} (h) \, \Sigma
(k)} {2 \sqrt{h^2 + |\Sigma (k)|^2}}
\left[ \langle n_{k \uparrow} \rangle_0 - \langle n_{k \downarrow}
\rangle_0 \right] \nonumber \\
&=& -\frac{\Sigma (k)} {2 h} \left[ \langle n_{k \uparrow} \rangle_0
- \langle n_{k \downarrow} \rangle_0 \right]. \label{H2}
\end{eqnarray}
Finally, from a comparison of Eqs.~\eqref{H0} and \eqref{H2}, the
self-energy in the limit of $T \ll |h|$ becomes
\begin{equation}
\Sigma (k) = 2|h| \tilde\alpha \Lambda \tilde\gamma (k).
\label{eq:4}
\end{equation}
The real and imaginary parts of $\Sigma (k)$ are plotted in
Fig.~\ref{fig:fig8}. Note that the complex phase of $\Sigma (k)
\propto \tilde\gamma (k)$ winds by $4\pi$ around the $\Gamma$ point.

\subsubsection{Berry curvature and Hall conductivity}
\label{sec:berry-curvature}

\begin{figure}[t]
\begin{center}
{\bf a)} \hspace{-3mm}\includegraphics[scale=0.30]{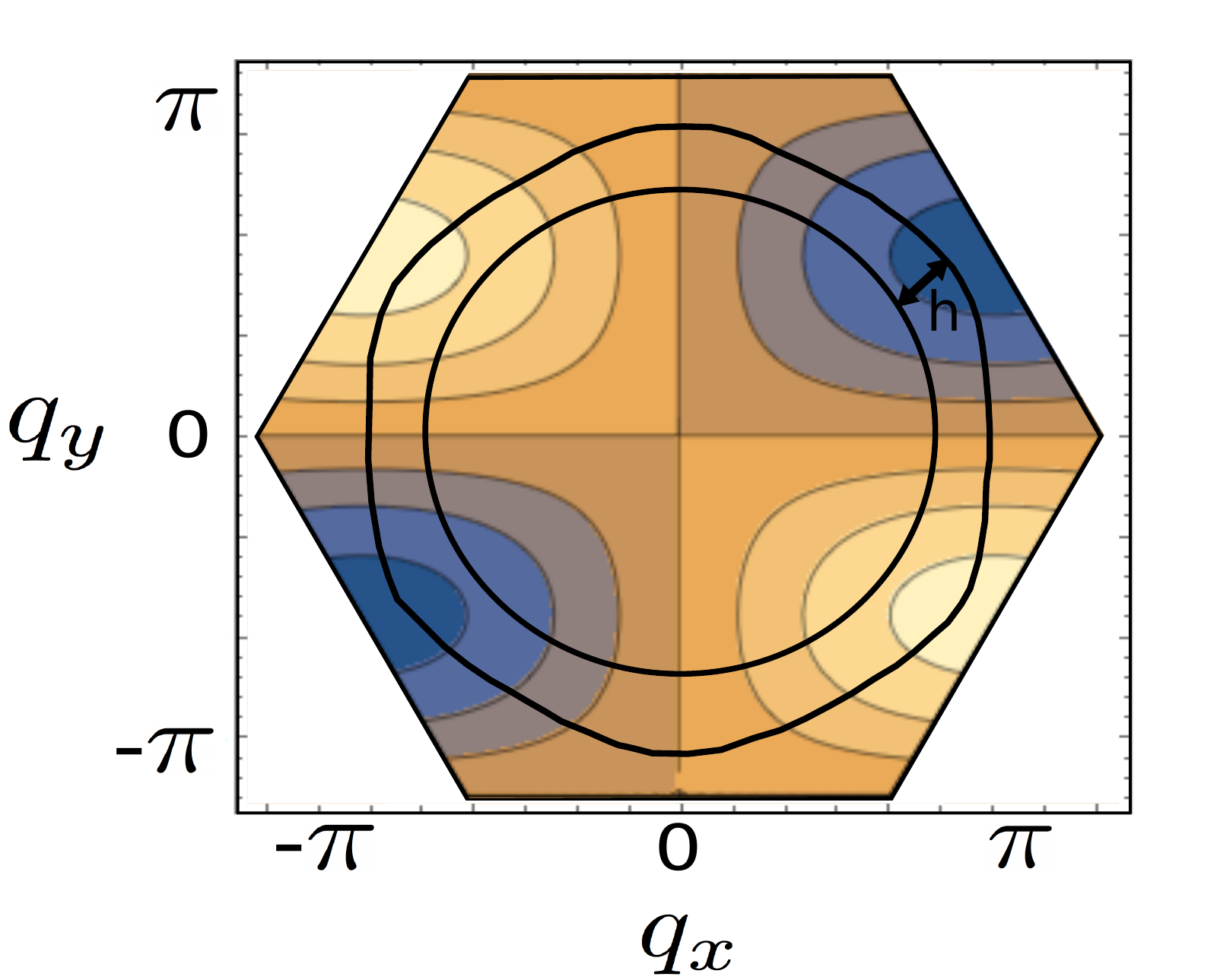}\,
\hspace{15mm} {\bf b)}
\hspace{-1mm}\includegraphics[scale=0.30]{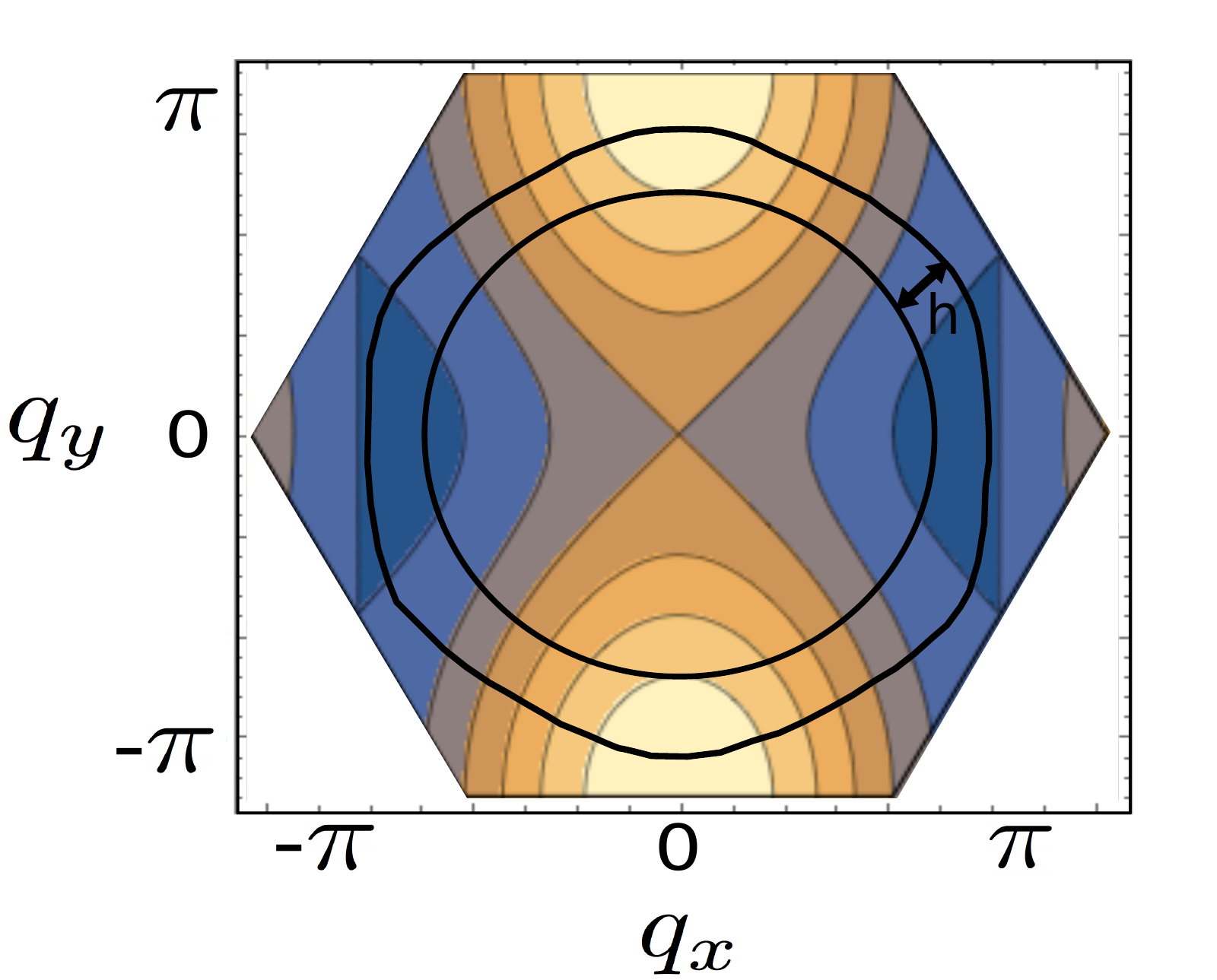}
\caption{The {\bf a)} real and {\bf b)} imaginary components of
$\Sigma(k)$ in a magnetic field $\vec{B} = h \hat{z}$.
     Lighter (darker) contours are positive (negative) contributions.
     The positions of the spin up and spin down Fermi surfaces in the presence of a nonzero Zeeman field $\propto h$ are also shown.
      }
\label{fig:fig8}
\end{center}
\end{figure}

Now we are in a position to calculate the emergent Hall
conductivity. First, we rewrite the effective quasiparticle
Hamiltonian in Eq.~(\ref{eq:2}) into the standard form
\begin{eqnarray}
\mathcal{H}_{\text{eff}} (k) &=& \left( f^\dagger_{k \uparrow} \,
f^\dagger_{k \downarrow} \right) \left[ \varepsilon (k) \sigma_0 - h
\vec{\beta} (k) \cdot \vec{\sigma} \right] \left(
\begin{array}{c} \f_{k \uparrow} \\ \f_{k \downarrow}
\end{array} \right), \nonumber \\
\vec{\beta} (k) &=& \left( -\frac{\mathrm{Re} \, \Sigma (k)} {h}, \,
-\frac{\mathrm{Im} \, \Sigma (k)} {h}, \, 1 \right), \label{eq:H-1}
\end{eqnarray}
where $| \vec{\beta} (k) | \approx 1$ in the limit of $| \Sigma (k)
| \ll |h|$. For such a Hamiltonian, the two bands have Berry
curvatures of opposite sign and equal magnitude given by
\begin{eqnarray}
\mathcal{B} (k) &\sim& \vec{\beta} (k) \cdot \big[
\partial_{k_x} \vec{\beta} (k) \times \partial_{k_y} \vec{\beta}
(k) \big] \nonumber \\
&\sim& \frac{1} {\rho_k} \big\{ \vec{\beta} (k) \cdot \big[
\partial_{\rho_k} \vec{\beta} (k) \times \partial_{\varphi_k} \vec{\beta}
(k) \big] \big\} \label{eq:B-1} \\
&\sim& \frac{1} {h^2 \rho_k} \, \mathrm{Im} \big[ \partial_{\rho_k}
\Sigma^* (k) \, \partial_{\varphi_k} \Sigma (k) \big], \nonumber
\end{eqnarray}
where we use polar coordinates defined by $k_x = \rho_k \cos
\varphi_k$ and $k_y = \rho_k \sin \varphi_k$. Due to the $4\pi$
phase winding of $\Sigma (k)$ (see Fig.~\ref{fig:fig8}), there is a
finite azimuthal derivative $\partial_{\varphi_k} \Sigma (k) \sim i
\Sigma (k)$. From $\partial_{\rho_k} \Sigma^* (k) \sim a \Sigma^*
(k)$, the Berry curvature at radius $\rho_k \sim 1/a$ is then on the
order of
\begin{equation}
\mathcal{B} (k) \sim \frac{a^2 | \Sigma (k) |^2} {h^2} \sim
\tilde\alpha^2 a^2 \left( \frac{h} {J_0} \right)^2. \label{eq:B-2}
\end{equation}
Next, in terms of the Berry curvatures $\pm \mathcal{B} (k)$ of the
two bands, the emergent Hall conductivity takes the form
\begin{equation}
\sigma^{qp}_{xy} = \int d^2 k \, \mathcal{B} (k) \left[ \langle n_{k
\uparrow} \rangle_0 - \langle n_{k \downarrow} \rangle_0 \right].
\end{equation}
In the limit of $T \ll |h|$, the integrand is only non-zero in an
annulus of thickness $\sim h / (a J_0)$ around the Fermi surface of
radius $\sim 1/a$, and the Hall conductivity can then be estimated
as $\sigma^{qp}_{xy} \sim \tilde\alpha^2 (h / J_0)^3$.

Finally, by virtue of the Wiedemann-Franz law that relates the
emergent and the thermal conductivities, the quasiparticle
contribution to the thermal Hall conductivity is on the order of
\begin{equation}
\kappa_{xy} \sim T \sigma^{qp}_{xy} \sim \tilde\alpha^2 T \left(
\frac{h} {J_0} \right)^3 \sim \frac{T h^3 J_{\pm z}^2} {J_0^5}.
\label{eq:kappa}
\end{equation}
Interestingly, $\kappa_{xy}$ is proportional to the third power of
the magnetic field. Note, however, that this result is valid for a
relatively large field ($T \ll |h| \ll J_0$). For a small field
($|h| \ll T \ll J_0$), the factor $\coth (h / T)$ in Eq.~(\ref{H0})
contributes an additional factor $\sim (T / h)^2$ to $\kappa_{xy}$,
which is then linearly proportional to the magnetic field.

\section{Discussion}

\subsection{Relationship to other Work}

In this paper, we have provided a comprehensive commentary on the
possibility of spin liquid physics in a very general spin-orbit
coupled model on the triangular lattice.
In the process, we have attempted to consolidate several previous
results on this topic. We began by looking at the $U(1)$ PSG wave
functions derived in Ref.~\cite{2016arXiv161203447L}. Instead of
working with these wave functions phenomenologically, we go beyond
their simple mean-field analysis and find quantitative estimates of
the energies of these ans\"atze using variational Monte Carlo.

We also use VMC to give a complete picture of magnetic order in our
model. Our results improve on the classical magnetic phase diagrams
presented in Refs. \cite{PhysRevB.94.035107, PhysRevB.94.174424}. In
those works, a phase transition between the 120$^\circ$ and stripe
phases is found in the nearest-neighbor model, and it is conjectured
that large spin fluctuations may lead to the presence of a
nonmagnetic phase.  In our work, by building on the PSG ans\"atze,
we also find a phase transition between the two magnetic phases in a
similar parameter regime. We further find that second-neighbor
interactions are necessary to create a spin liquid ground state and
we identify the Dirac spin liquid as the lowest energy state.  This
confirms and extends earlier studies of the isotropic Heisenberg
model \cite{PhysRevB.93.144411}.

The only other calculation of the full quantum phase diagram in this
model was given by the DMRG analysis in
Ref.~\cite{2017arXiv170302971Z}. Our phase diagram agrees with the
DMRG analysis when second-neighbor interactions are included.  The
XXZ anisotropy and $J_{\pm\pm}$ interactions both work to limit the
spin liquid phase to a very small region of parameter space.
However, we go beyond this and also include a third-neighbor
interaction, which we believe gives a more complete picture on the
behavior of the spin liquid phase.  We find that even a very small
third-neighbor interaction can greatly stabilize the spin liquid
regime.

\subsection{Relevance to Materials}

This model has recently attracted much attention for its potential
relevance to the material YbMgGaO$_4$.  Experiments find enticing
evidence for a spinon Fermi surface state from thermodynamic and
inelastic neutron scattering measurements \cite{nature,natphys}. Our
work addressed the theoretical basis for such physics.

Our results support the claim of Ref.~\cite{2017arXiv170302971Z}
that YbMgGaO$_4$ likely falls outside of the spin liquid phase in
the presence of only first- and second-neighbor interactions. We
found, however, that a very small third-neighbor interaction can
greatly increase the size of the spin liquid phase and may appear
quite naturally in the material. However, using the simple PSG
picture, we always find that the Dirac spin liquid is energetically
favored over the spinon Fermi surface state.

While the above results do not support the spinon Fermi surface
state, we did find some effects which could tilt the balance in its
favor.  We saw that the spin-orbit interactions favor the spinon
Fermi surface over the Dirac spin liquid state when we include
effects beyond the simple projected mean-field wave functions.  This
leaves open the possibility that the spinon metal could be
energetically favorable, perhaps assisted by other factors such as
disorder or a small ring-exchange interaction.

If we assume that a spinon metal state does exist, interesting
features emerge due to spin-orbit coupling. We showed how the
spin-orbit interactions could explain the existence of broad peaks
at the $M$ points in the spin structure factor and also predicted
that measurements of the spin-polarized structure factors would
display specific polarization-dependent peaks reflecting the
anisotropic interactions.  We also propose that the spin-orbit
coupled spinon metal state may have a rather large thermal Hall
conductivity which could be a very clear signature of spin liquid
physics in such a system.

\subsection{Future directions and implications}
\label{sec:future-direct-impl}

Looking forward, we anticipate a number of implications for the
results and techniques developed in this work.  For our spin-orbit
coupled triangular systems, we showed that the restrictions imposed
on the standard Gutzwiller-projected free fermion states by the PSG
are quite severe for several of the $U(1)$ QSL states. Consequently,
they are unable to adapt to strongly anisotropic interactions, and
this may open the door to competition from $\mathbb{Z}_2$ QSL states
in the case of such anisotropic models. In turn, this would be of
considerable interest as the Gutzwiller-based approach almost always
favors $U(1)$ states in Heisenberg models. The possibility of
inducing fully gapped topological QSLs should be explored in the
future by VMC techniques.

We argued that the thermal Hall effect should be a key signature of
itinerant spinon excitations in spin-orbit coupled systems.  While
we obtained such an effect for the $U(1)$ spinon Fermi surface state
on the triangular lattice, it was in fact suppressed by the
PSG-mandated vanishing of effective spin-orbit coupling on the
fermionic spinons at the free-particle level.  Ultimately, this
suppression owes itself to the presence of inversion symmetry,
which, in conjunction with time-reversal symmetry, act on the
spinons analogously to the way they do on real electrons.  As is
well known, the combination of inversion and time reversal in that
context imply an exact two-fold Kramers degeneracy of the full
electronic band structure, and a similar effect occurs here.  When
inversion is absent, for example, when an electric field is present
normal to a two-dimensional electron gas, spin splitting occurs. The
Rashba spin-orbit coupling induced by such a field is known to
induce a large anomalous Hall effect in that context
\cite{nagaosa2010anomalous}.  This strongly suggests that one should
look for an enhanced thermal Hall effect in two-dimensional magnetic
materials in which the magnetic layer has an asymmetric environment.
This criteria, along with the requirement of large spin-orbit
coupling, should assist in a search for this phenomenon.

Our methodology offers a consistent and quantitative method to
compare QSLs and ordered phases for anisotropic magnetic
Hamiltonians.  This should have broad applicability to other
materials such as the Kitaev compounds $\alpha$-RuCl$_3$,
Na$_2$IrO$_3$, and Li$_2$IrO$_3$ in all its structural variations,
and to three-dimensional systems like rare earth pyrochlores and
spinels.  The ability of VMC-based methods to tackle large systems
is a unique numerical advantage.  We expect many insights from such
studies in the future.

\section*{Acknowledgments}
We thank Radu Coldea, Martin Mourigal, Gang Chen, and Yuan-Ming Lu
for useful discussions.  This work was supported by the NSF
Materials Theory program through Grant No.~DMR1506119 (J.I., C.L.,
L.B.) and by the Gordon and Betty Moore Foundation's EPiQS
Initiative through Grant No.~GBMF4304 (G.B.H.).
We acknowledge support from the Center for Scientific Computing from the CNSI,MRL: an NSF MRSEC (DMR-1121053).

\bibliography{base}

\end{document}